\documentclass[iop, numberedappendix]{mn2e}

\usepackage[pdftex]{graphicx}

\usepackage{amsmath, amsthm, amssymb}
\let\bibhang\relax
\usepackage{natbib}
\usepackage{subfigure}
\usepackage{upgreek}
\usepackage{accents}
\usepackage{color}
\usepackage{hyperref}
\usepackage{dblfloatfix}
\usepackage[T1]{fontenc}
\usepackage{aecompl}
\usepackage{enumitem}
\newcommand{\angstrom}{\text{\normalfont\AA}}

\hypersetup{
  colorlinks   = true, %Colours links instead of ugly boxes
  urlcolor     = blue, %Colour for external hyperlinks
  linkcolor    = blue, %Colour of internal links
  citecolor   = blue %Colour of citations
}

\newcommand{\Rvir}{R_{\text{vir}}}

\newcommand{\sigmaN}{\sigma_{\log_{10} N_d}}

\newcommand{\univ}{\texttt{univ}}
\newcommand{\evol}{\texttt{evol}}
\newcommand{\evole}{\texttt{evol-e}}
\newcommand{\evolm}{\texttt{evol-m}}

\newcommand{\rvir}{R_{\text{vir}}}

\newcommand{\alphalo}{\alpha_{\ast,\text{lo}}}
\newcommand{\alphahi}{\alpha_{\ast,\text{hi}}}

\newcommand{\MUV}{M_{\text{UV}}}

\newcommand{\Mpeak}{M_{\text{peak}}}

\newcommand{\SFR}{\dot{M}_{\ast}}

\newcommand{\bMAR}{\dot{M}_b}

\newcommand{\fduty}{f_{\text{duty}}}

\newcommand{\fdtmr}{f_{\text{dtmr}}}

\newcommand{\fdutynorm}{f_{\text{duty,10}}}
\newcommand{\fdutyalpha}{\alpha_{\text{duty}}}
\newcommand{\fdutygamma}{\gamma_{\text{duty}}}

\newcommand{\fdtmrnorm}{f_{\text{dtmr,10}}}
\newcommand{\fdtmralpha}{\alpha_{\text{dtmr}}}
\newcommand{\fdtmrgamma}{\gamma_{\text{dtmr}}}

\newcommand{\fstar}{f_{\ast}}

\newcommand{\mmin}{m_{\text{min}}}
\newcommand{\mmax}{m_{\text{max}}}

\newcommand{\Msun}{M_{\odot}}

\title[Distinguishing high-$z$ galaxy evolution scenarios]{Prospects for distinguishing galaxy evolution models with surveys at redshifts $z \gtrsim 4$}
\author[J. Mirocha]{
Jordan Mirocha\textsuperscript{\thanks{jordan.mirocha@mcgill.ca}}\textsuperscript{\thanks{CITA National Fellow}} \\
McGill University Department of Physics \& McGill Space Institute, 3600 Rue University, Montr\'eal, QC, H3A 2T8 \\
}

\begin{document}

\pagerange{\pageref{firstpage}--\pageref{lastpage}} \pubyear{2020}
\maketitle

%%%
%% Abstract
%%%
\begin{abstract}
Many semi-empirical galaxy formation models have recently emerged to interpret high-$z$ galaxy luminosity functions and make predictions for future galaxy surveys. A common approach assumes a ``universal'' star formation efficiency, $f_{\ast}$, independent of cosmic time but strongly dependent on the masses of dark matter halos. Though this class of models has been very successful in matching observations over much of cosmic history, simple stellar feedback models \textit{do} predict redshift evolution in $f_{\ast}$, and are commonly used in semi-analytic models. In this work, we calibrate a set of universal $f_{\ast}$ and feedback-regulated models to the same set of rest-ultraviolet $z \gtrsim 4$ observations, and find that a rapid, $\sim (1+z)^{-3/2}$ decline in \textit{both} the efficiency of dust production \textit{and} duty cycle of star formation are needed to reconcile feedback-regulated models with current observations. By construction, these models remain nearly identical to universal $f_{\ast}$ models in rest-ultraviolet luminosity functions and colours. As a result, the only way to distinguish these competing scenarios is either via (i) improved constraints on the clustering of galaxies -- universal and feedback-regulated models differ in predictions for the galaxy bias by $0.1 \lesssim \Delta \langle b \rangle \lesssim 0.3$ over $4 \lesssim z \lesssim 10$ -- or (ii) independent constraints on the dust contents and/or duty cycle of star formation. This suggests that improved constraints on the `dustiness' and `burstiness' of high-$z$ galaxies will not merely add clarity to a given model of star formation in high-$z$ galaxies, but rather fundamentally determine our ability to identify the correct model in the first place.
\end{abstract}

%%%
%% Introduction
%%%
\section{Introduction}
The frontier in observations of galaxies has continued its march to higher redshifts in recent years, with substantial samples ($N \sim 10^4$) now established at $4 \lesssim z \lesssim 8$ \citep{Bouwens2015,Finkelstein2015}, with growing numbers even at $8 \lesssim z \lesssim 12$ \citep[e.g.,][]{Oesch2018,Bowler2020,Morishita2018,McLeod2016,Livermore2018,RojasRuiz2020,Stefanon2019}. Rest-ultraviolet (UV) measurements with the Wide-Field Camera 3 (WFC3) on the \textit{Hubble Space Telescope} (HST) are largely to thank for this progress, and soon observations with the \textit{James Webb Space Telescope} (JWST) will continue along the path forged by many successful \textit{Hubble} programs \citep{Windhorst2011,Koekemoer2011,Grogin2011,Bouwens2011,Illingworth2013}.

The assembly of statistical samples of high-$z$ galaxies has spawned a number of independent efforts to understand the shape and redshift evolution of the galaxy luminosity function (LF) in the context of structure formation in a $\Lambda$CDM cosmology. With basic ingredients like the dark matter (DM) halo mass function (HMF), halo mass accretion rate (MAR), or perhaps merger trees drawn from N-body simulations, many groups have shown that evolution in the high-$z$ galaxy population can be described largely by evolution in the dark matter (DM) halo population, assuming a tight link between galaxy and halo growth \citep[e.g.,][]{Trenti2010,Behroozi2013,Dayal2013,Tacchella2013,Mason2015,Sun2016,Tacchella2018,Yung2019a,Behroozi2019}. Tuning of these models \textit{is} required, though the calibration of models to LFs at a single redshift generally results in predictions that agree well with measurements even at redshifts neglected in the calibration. As a result, predictions for the galaxy LF at redshifts and/or magnitudes beyond current detection limits rest on a strong empirical foundation \citep[see also, e.g.,][]{Williams2018}.

Of course, the success of empirically-calibrated models does not necessarily imply that we understand galaxy evolution at high redshift. For example, pure ``semi-empirical'' models, which parameterize the efficiency with which stars form and calibrate its free parameters empirically, generally find that high-$z$ LFs can be fit under the assumption of a mass-dependent but time-\textit{independent} star formation efficiency (SFE). Though the inferred mass-dependence of this relation is roughly consistent in many cases with the arguments of stellar feedback models, which predict $f_{\ast} \propto M_h^{1/3-2/3}$ \citep{Dayal2013,Furlanetto2017}, these same stellar feedback models also predict time evolution in $f_{\ast}$, at the level of $f_{\ast} \propto (1+z)^{1/2-1}$ at fixed halo mass \citep[see also, e.g.,][]{Murray2005}. Does the perceived lack of evolution indicate a departure from feedback models, and thus some new insight into how star formation proceeds in early galaxies, or is it instead a symptom of the failure of another component of the model, e.g., dust obscuration, and/or the assumption of smooth, inflow-driven star formation?

Rather than flexibly parameterizing the SFE and fitting for its parameters, some semi-analytic models (SAMs) impose the SFE \textit{a priori}, and use other parameters to reduce tension between model and data. For example, \citet{Somerville2012} noted that their SAM required an evolving dust optical depth in order to avoid under-producing the number density of massive, UV bright galaxies at high redshift. Because the \citet{Somerville2012} SAM assumes star formation is regulated by  energy-driven winds and is thus redshift-dependent, evolution in the dust law or dust geometry seems a plausible way to counteract evolution in the SFE. Similar adaptations have been employed in other recent works \citep[e.g.,][]{Yung2019a,Vogelsberger2019,Qiu2019}. However, it is not clear that this is the only way to reconcile physically-motivated models of star formation with observations, or if such scenarios can be differentiated from simpler models with no evolution in the SFE or dust production, like those put forth recently in \citet{Mirocha2020}.

The goal of this paper is to explore several ways in which physically-motivated models of star formation can be reconciled with current measurements at $z \gtrsim 4$, and to determine if these scenarios can be distinguished from the common semi-empirical approach that assumes a universal SFE. In addition to the possibility of evolution in the properties of dust, we also allow burstiness in the star formation histories of galaxies. In each case, we allow quantities of interest to depend both on halo mass and redshift, rather than invoking redshift-dependent correction factors that operate on all halos identically.

In Section \ref{sec:model}, we review the basic components of our model, which has been described in large part elsewhere. In Sections \ref{sec:results} and \ref{sec:discussion} we present our main results and offer some discussion in the context of ongoing work in the community, respectively. We conclude in Section \ref{sec:conclusions}.

We adopt AB magnitudes throughout \citep{Oke1983}, i.e.,
\begin{equation}
    M_{\lambda} = -2.5 \log_{10} \left(\frac{f_{\lambda}}{3631 \ \mathrm{Jy}} \right)
\end{equation}
and adopt the following cosmology: $\Omega_m = 0.3156$, $\Omega_b = 0.0491$, $h = 0.6726$, and $\sigma_8=0.8159$, very similar to the recent \citet{Planck2018} constraints.

%%
% MODEL
%%
\section{Model} \label{sec:model}
Our approach to modeling high-$z$ galaxies is similar to other models in the literature \citep[e.g.,][]{Sun2016,Mason2015}, has been described in large part in earlier papers \citep[e.g.,][]{Mirocha2017}, and is publicly available within the \textsc{ares}\footnote{\url{https://ares.readthedocs.io/en/latest/}} code. The components most pertinent to the present paper are presented in \citet{Mirocha2020}, to which we refer the reader for quantitative details, particularly with regards to the modeling of UV colours.

Briefly, we model the SFE as a double-power law in halo mass,
\begin{equation}
    \fstar(M_h) = \frac{f_{\ast,10} \ \mathcal{C}_{10}} {\left(\frac{M_h}{M_{\mathrm{p}}} \right)^{-\alphalo} + \left(\frac{M_h}{M_{\mathrm{p}}} \right)^{-\alphahi}} \label{eq:sfe_dpl}
\end{equation}
where $f_{\ast,10}$ is the SFE at $10^{10} M_{\odot}$, $M_p$ is the mass at which $\fstar$ peaks, and $\alphahi$ and $\alphalo$ describe the power-law index at masses above and below the peak, respectively. The additional constant $\mathcal{C}_{10} \equiv (10^{10} / M_p)^{-\alphalo} + (10^{10} / M_p)^{-\alphahi}$ is introduced to re-normalize the standard DPL formula to $10^{10} M_{\odot}$, rather than the peak mass. In this standard form, the model is ``universal'' in that no component of $f_{\ast}$ is allowed to vary with redshift, and is thus labeled as our \texttt{univ} model throughout.

The star formation histories of galaxies are generated assuming that the star formation rate (SFR) is simply $\dot{M}_{\ast} = f_{\ast} \dot{M}_b$, where $\dot{M}_b$ is the baryonic MAR of the galaxy. We assume $\dot{M}_b$ is the total MAR times the cosmic baryon fration, $f_b$, i.e., $\dot{M}_b= f_b \dot{M}_h$, and we generate $\dot{M}_h$ from the HMF itself assuming halos grow at fixed abundance \citep[see Appendix A of][]{Furlanetto2017}. We further assume 0.3 dex scatter in SFR at fixed halo mass, and spawn 10 halos in each mass bin (of width $\Delta \log_{10} M_h = 0.01$), thinning the number density of halos in each bin accordingly to preserve the overall abundance. We assume that galaxies occupy halos in a 1:1 fashion and that the HMF is that given by \citet{Tinker2010}, which we compute using the \textsc{hmf}\footnote{\url{https://hmf.readthedocs.io/en/latest/}} code \citep{Murray2013}.

We further assume that the metal production rate, $\dot{M}_Z$, is proportional to the SFR, and that the dust yield is a fraction $\fdtmr$\footnote{Note that this is equivalent to the parameter $f_d$ in \citet{Mirocha2020}. We change the notation here to avoid confusion with the duty cycle, $\fduty$.} of the metal yield. Assuming an effective dust scale length, $R_d$, the dust optical depth is thus given by
\begin{equation}
    \tau_{\nu} = \kappa_{\lambda} N_d = \kappa_{\lambda} \frac{3 \fdtmr M_Z}{4 \pi R_d^2} \label{eq:tau_d}
\end{equation}
where $\kappa \propto \lambda^{-1}$ is the dust opacity. We perform spectral synthesis over the entire star formation history (SFH) of each galaxy in the model, adopting the \textsc{bpass} version 1.0 models \citep{Eldridge2009}, reddened by an optical depth given by Eq. \ref{eq:tau_d}, and ``observe'' galaxies using the relevant HST filters as a function of redshift \citep[in accordance with][]{Bouwens2014}. In \citet{Mirocha2020}, we found that this model naturally fits UVLFs and generates mild redshift evolution in the $\MUV$-$\beta$ relation, consistent with observations, despite no underlying evolution in dust properties. See \S2 in \citet{Mirocha2020} for more details.

In what follows, we add two more models representative of energy- and momentum-regulated feedback, which we refer to as \evole\ and \evolm\, due to the evolution they induce in the SFE. The basic changes relative to the \univ\ model are as follows:
\begin{itemize}
	\item While we still adopt a double power-law model for the SFE, we force the low-mass behavior of $f_{\ast}$ to follow expectations for energy- and momentum-regulated stellar feedback, i.e., with $f_{\ast,10} \propto (1 + z)$ (or $(1+z)
  ^{1/2}$) and $\alphalo=2/3$ (or $\alphalo=1/3$) for energy (or momentum) regulated feedback \citep{Furlanetto2017}. These trends emerge when one assumes that star formation occurs until (i) the rate of energy injection from supernovae is equal to the rate at which accreting gas accumulates binding energy, or (ii) the momentum imparted to the gas (again, presumably by supernova blastwaves) is able to accelerate halo gas to the escape velocity. In reality, a mixture of these feedback mechanisms is likely at work, though each is a useful limiting case -- a point which we revisit in \S\ref{sec:implications_feedback}. For each feedback model, we still allow the mass at which $f_{\ast}$ peaks, $\Mpeak$, and the slope in $f_{\ast}$ above the peak, $\alphahi$, to vary as free parameters.
	\item We also adopt a double power-law model for the dust scale length, $R_d$, but force the mass and redshift-dependence of the high-mass component to evolve like the virial radius of DM halos, i.e., $R_d \propto M_h^{1/3} (1+z)^{-1}$. This is to remain conceptually consistent with the feedback-regulated star formation models, which are also driven by the size evolution of halos (through the halo binding energy and escape velocity). The normalization of this relationship, its pivot mass, and low-mass slope are still allowed to vary freely.
\end{itemize}
These choices introduce a significant amount of tension between model predictions and observational constraints.

\begin{figure*}
\begin{center}
\includegraphics[width=0.98\textwidth]{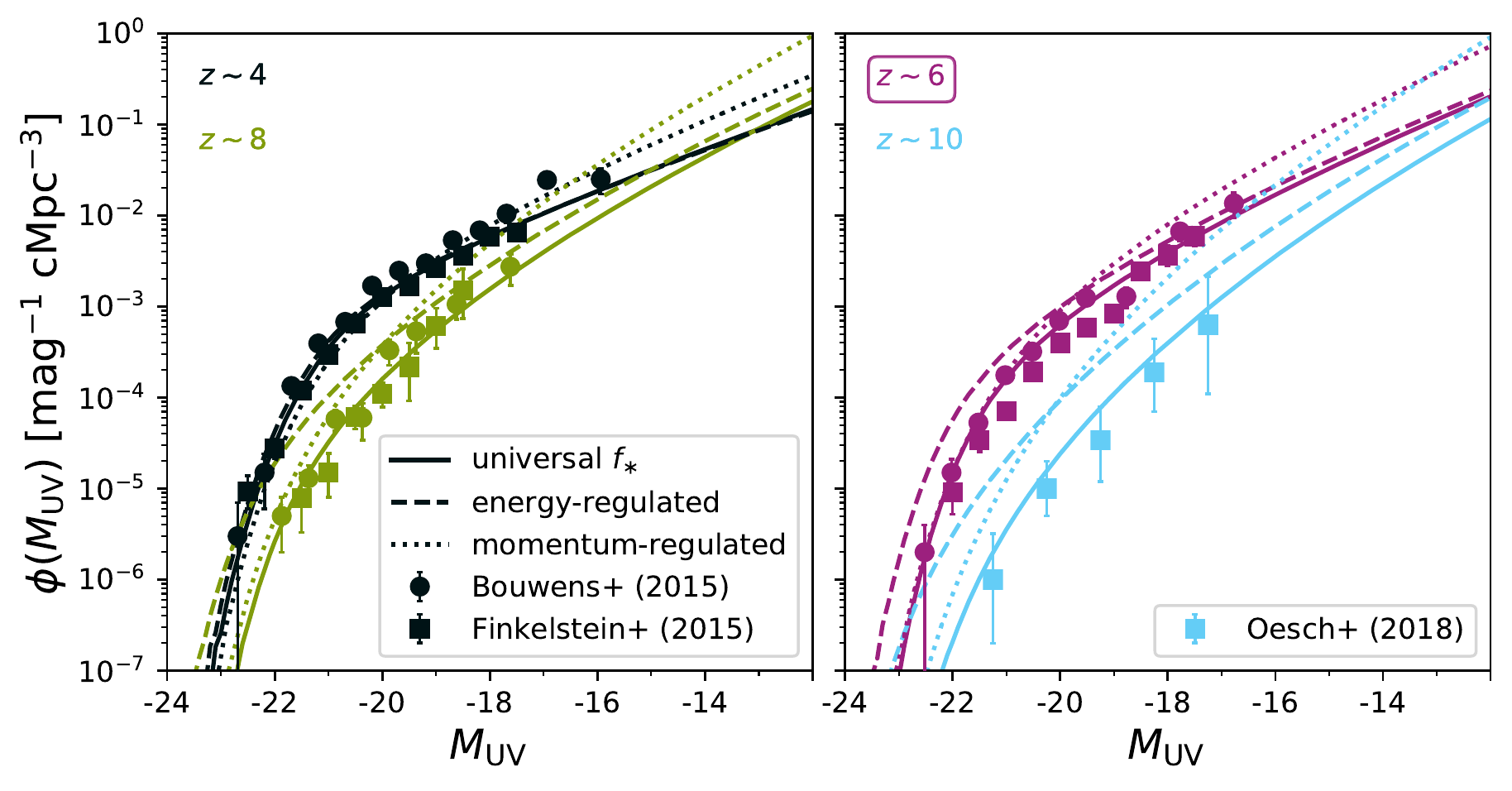}
\caption{{\bf Evolution in $f_{\ast}$ quickly produces tension in high-$z$ UVLFs.} UVLFs at $z=4,6,8,$ and 10 are shown compared to the \citet{Bouwens2015} and \citet{Finkelstein2015} measurements at $4 \lesssim z \lesssim 8$ and \citet{Oesch2018} at $z \sim 10$. Solid lines in each panel indicate the universal SFE solution, calibrated to the $z \sim 6$ data from \citet{Bouwens2015}, while dashed (dotted) lines indicate predictions of energy (momentum) regulated feedback models. Note that $f_{\ast}$ is normalized such that each model is comparable at $z=4$, and each model neglects dust obscuration for clarity.}
\label{fig:zevol_sfe}
\end{center}
\end{figure*}

In Figure \ref{fig:zevol_sfe}, we illustrate first the effects of SFE evolution. For simplicity, we ignore dust to isolate the evolving SFE effects, which means these models would not produce viable $\MUV$-$\beta$ relations. We will add dust back into the model momentarily. With the UVLF at $z = 4$ roughly preserved for all models, we see clearly that energy-regulated (dashed) and momentum-regulated (dotted) models cause a systematic over-prediction of the luminosity of $z > 4$ galaxies. This trend is apparent when comparing to either the \citet{Bouwens2015} (cirles) and \citet{Finkelstein2015} (squares) UVLFs. One solution to this problem would be to invoke a rise in dust attenuation with redshift, though dust of course affects UV colours as well, as we describe next.

\begin{figure*}
\begin{center}
\includegraphics[width=0.98\textwidth]{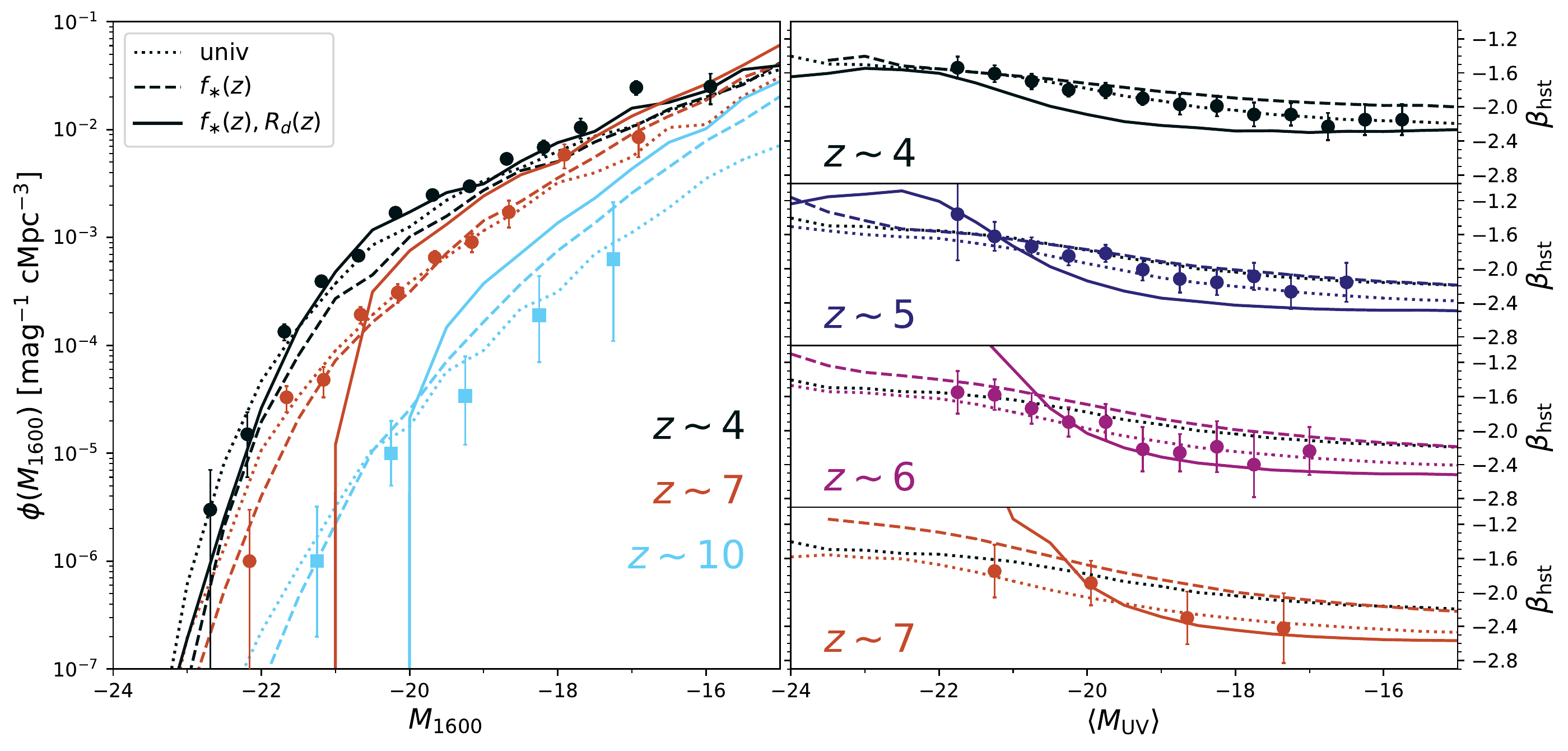}
\caption{{\bf Evolution in the SFE and dust scale length cause substantial tension with observed UVLFs (left) and $\MUV$-$\beta$ relations (right).} Dotted lines indicate the \texttt{univ} model from \citet{Mirocha2020}, while dashed lines are a model identical to the \texttt{univ} model but with $(1+z)$ evolution in the SFE, which reverses the redshift evolution in UV colours at fixed $\MUV$ and worsens agreement with UVLFs. Solid lines augment the model further, forcing $f_{\ast} \propto (1+z)$ and $R_d \propto \Rvir \propto M_h^{1/3} (1+z)^{-1}$. In this case, the rapid decline in dust scale length with redshift (at fixed halo mass) results in dramatic over-reddening of galaxies. For illustrative purposes here, we increase the normalization of the dust scale length by a factor of $3$ relative to the \texttt{univ} model to mitigate catastrophic over-reddening of the massive galaxy population. Note that the $z\sim 4$ $\MUV$-$\beta$ relation of the \texttt{univ} model (dotted black) is repeated in each panel to guide the eye.}
\label{fig:zevol_Rd}
\end{center}
\end{figure*}

Without dust, the UV colours of model galaxies will be uniformly blue, and thus in disagreement with constraints on the $\MUV$-$\beta$ relation at $4 \lesssim z \lesssim 7$, which indicate non-negligible reddening and evolution in $\beta$ with UV magnitude at a significant level $\Delta \beta \sim 1$ over $-22 \lesssim \MUV \lesssim -16$ \citep{Finkelstein2012,Bouwens2014}. Such models can still be tuned to match the bright-end of UVLFs by varying the efficiency of star formation in high-mass halos, as is the case in Figure \ref{fig:zevol_sfe}. In Figure \ref{fig:zevol_Rd}, we compare the \texttt{univ} model of \citet{Mirocha2020} (dotted) to two different extensions, which include evolution in the SFE (dashed), and both the SFE and dust scale length (solid). Evolution in $f_{\ast}$ worsens agreement with UVLFs, and causes UV colours to get redder with increasing redshift (at fixed $\MUV$) -- the opposite of the observed trend. To make matters worse, if dust scale lengths shrink rapidly as $R_d \propto \Rvir \propto M_h^{1/3} (1+z)^{-1}$, the reddening grows more extreme, the bright end of the UVLF and $\MUV$-$\beta$ relation become much too steep (solid lines). As shown in \citet{Mirocha2020}, scatter in the dust column density (at fixed $M_h$) can help bridge the gap between these extremes, though cannot obviously fix UVLFs and $\MUV$-$\beta$ relations simultaneously at all redshifts.

To alleviate the tensions illustrated by Figures \ref{fig:zevol_sfe} and \ref{fig:zevol_Rd}, we introduce two new degrees of freedom in the model, which we described below.

First, we consider a non-unity duty cycle, which we model as a power-law in mass and redshift,
\begin{equation}
	\fduty = f_{\mathrm{duty},10} \left(\frac{M_h}{10^{10} \ M_{\odot}} \right)^{\alpha_{\mathrm{duty}}} \label{eq:fduty}
\end{equation}
where $f_{\mathrm{duty},10}$ is allowed to evolve with redshift as a power-law, and we impose a maximum of $\fduty = \mathrm{min(Eq. \ref{eq:fduty}}, 1)$. Whereas increasingly efficient star formation tends to result in overly-luminous galaxies, $\fduty < 1$ compensates by reducing the abundance of galaxies that are ``on'' at any given time. We impart this stochasticity randomly for each galaxy at each time-step by setting $\dot{M}_{\ast} = 0$ if a random number, $r$, drawn from a uniform distribution in the interval $\{0, 1\}$ is $r \geq \fduty$. In detail, $\fduty$ cannot be treated by a binary on/off approach given the non-trivial stellar aging effects at play, which we here treat self-consistently. As a result, $\fduty$ affects UVLFs and the $\MUV$-$\beta$ relation.

We also explore also the possibility that the efficiency with which dust is produced (or destroyed) depends on halo mass and/or redshift, which we do through the $\fdtmr$ parameter. Again, we adopt a power-law in mass,
\begin{equation}
	\fdtmr = f_{\mathrm{dtmr},10} \left(\frac{M_h}{10^{10} \ M_{\odot}} \right)^{\alpha_{\mathrm{dtmr}}} \label{eq:fdtmr}
\end{equation}
and allow power-law evolution in the normalization, i.e., $f_{\mathrm{dtmr},10}=f_{\mathrm{dtmr},10}(z)$. The effect of $\fdtmr$ variations is straightforward: a decline in the dust contents of galaxies renders their spectra bluer and UV luminosity higher.

As in \citet{Mirocha2020}, we use the affine-invariant Markov Chain Monte-Carlo (MCMC) code \textsc{emcee}\footnote{\href{https://emcee.readthedocs.io/en/stable/}{https://emcee.readthedocs.io/en/stable/}}\citep{ForemanMackey2013} to map the posterior distribution of the parameters given the data, which we take to be the UVLFs ($z \sim 4,6,$ and 8) and $\MUV$-$\beta$ relations ($z \sim 4$ and 6) from \citet{Bouwens2015} and \citet{Bouwens2014}, respectively.

%%
% RESULTS
%%
\section{Results} \label{sec:results}
Our goal in the remainder of the paper is to (i) determine the behaviour in $\fduty$ and/or $\fdtmr$ needed in order to reconcile energy-regulated feedback models with rest-UV data at high-$z$, and (ii) determine if either \texttt{evol} model has features that distinguish it from the \univ\ model of \citet{Mirocha2020}.

%\begin{figure*}
%\begin{center}
%\includegraphics[width=0.98\textwidth]{uvlf_cmd_eduty.pdf}
%\includegraphics[width=0.98\textwidth]{uvlf_cmd_edtmr.pdf}
%\caption{{\bf Evolution of UVLF and CMDs for models employing flexibility in either $\fduty$ (top) or $\fdtmr$ (bottom).} Left and center panels show UVLFs at a series of redshifts, while right-most panels show the evolution in the $\MUV$-$\beta$ relation over $4 \lesssim z \lesssim 7$. Filled contours indicate 68\% confidence reconstructions, while open contours indicate 95\%. Note the correspondence between 95\% confidence contours for UVLFs and $\MUV$-$\beta$ -- dashed (dotted) lines trace the faint (bright) edge of UVLF constraints and the faint/red (bright/blue) edge of $\MUV$-$\beta$ constraints, illustrating the tension in simultaneously fitting both datasets with single component additions to the model (either $\fduty$ \textit{or} $\fdtmr$, not both). All models assume energy-regulated stellar feedback and dust scale lengths that evolve like halo virial radii.}
%\label{fig:scam_failure}
%\end{center}
%\end{figure*}

\begin{figure*}
\begin{center}
\includegraphics[width=0.98\textwidth]{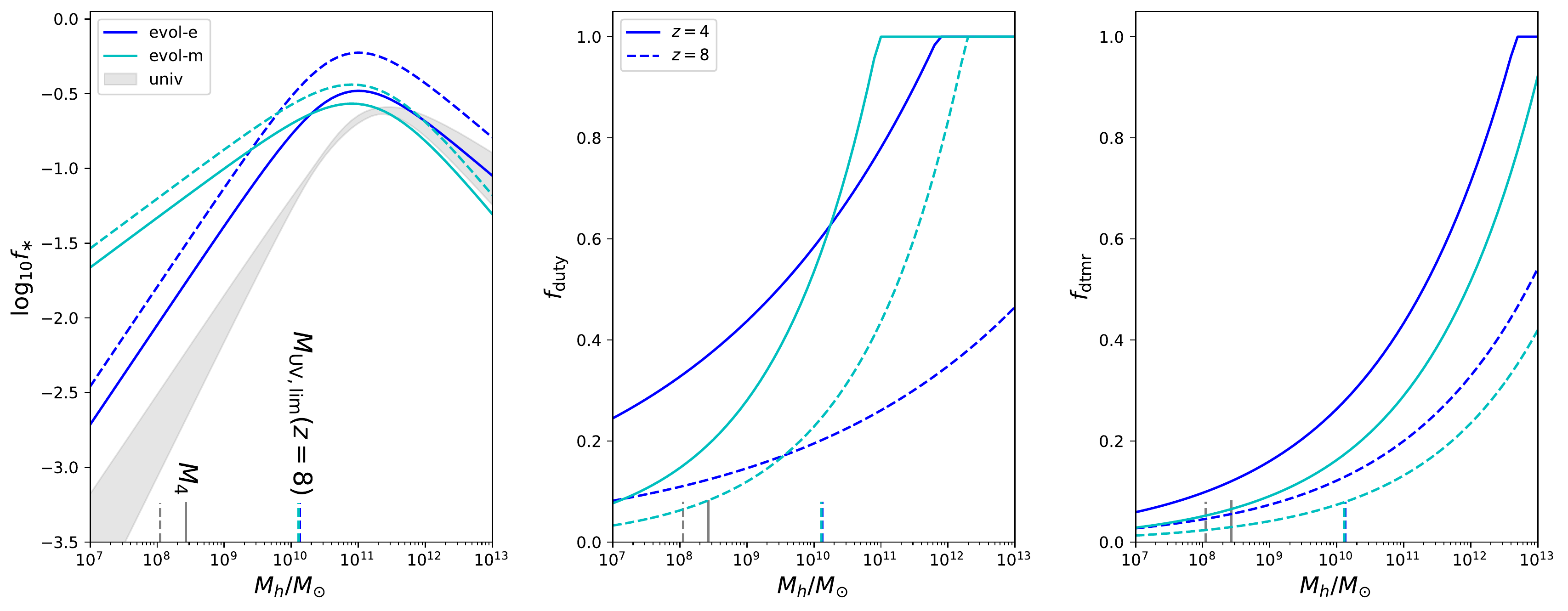}
\caption{{\bf Obtaining good fits to both UVLFs and $\MUV$-$\beta$ relations with our \texttt{evol} models require strong evolution in $\fduty$ and $\fdtmr$.} \textit{Left:} Star formation efficiency, $f_{\ast} \equiv \SFR/\bMAR$, of \texttt{univ} model (shaded region) and energy-regulated models (\evole) (blue) and momentum-regulated models (\evolm) (cyan) at $z=4$ and $8$ (solid and dashed, respectively). \textit{Middle:} Best fitting duty cycle for each model at $z=4$ and $z=8$. \textit{Right:} Best fitting $\fdtmr$ at $z = 4$ and $8$. In all panels, the atomic threshold at $z=4$ and $8$ is denoted along the bottom axis ($M_4$; gray), as is the halo mass corresponding to the limiting magnitude of current observations at $z \sim 8$, which is roughly $\MUV=-18$. Note that the $M_h$-dependence inferred for $\fduty$ and $\fdtmr$ is only mildly preferred in the energy-regulated model, as shown in detail in Figure \ref{fig:pdfs}, though the need for a redshift dependence is significant.}
\label{fig:recon}
\end{center}
\end{figure*}

\begin{figure*}
\begin{center}
\includegraphics[width=0.98\textwidth]{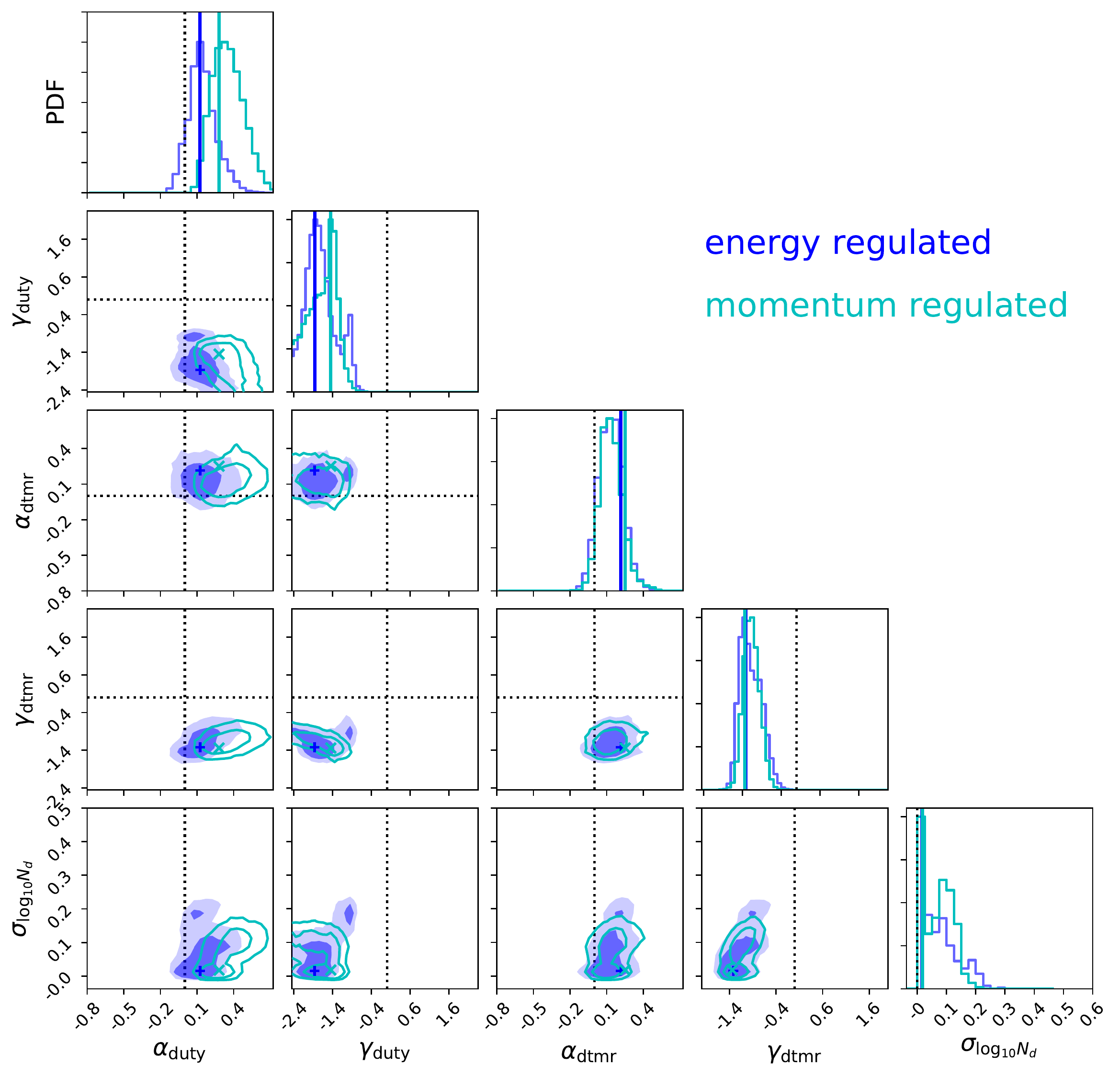}
\caption{{\bf Evolution in the SFE drives a need for evolution in both $\fduty$ and $\fdtmr$ at the level of $\sim (1+z)^{-3/2}$ or faster for both feedback models.} Here, we show the 68\% and 95\% confidence regions for parameters governing the power-law mass ($\alpha$) and redshift ($\gamma$) dependences of $\fduty$ and $\fdtmr$, for each feedback model (energy in blue, momentum in cyan). Dotted lines in each panel indicate the points in parameter space corresponding to no mass and/or redshift dependence. We also include the scatter in dust column density (at fixed $M_h$) in the first row, which is degenerate with both $\fduty$ and $\fdtmr$. Best-fits are indicated by solid vertical lines in panels along the diagonal and crosses in interior panels.}
\label{fig:pdfs}
\end{center}
\end{figure*}

First, in Figures \ref{fig:recon} and \ref{fig:pdfs}, we show the reconstructed $\fstar$, $\fduty$, and $\fdtmr$ curves obtained from the multi-dimensional fits and their posterior distributions of the most relevant parameters, respectively. While the reconstructed SFE (left panel of Fig. \ref{fig:recon}) is largely set by-hand, the behaviour of $\fduty$ and $\fdtmr$ arise in order to simultaneously match UVLFs and $\MUV$-$\beta$ relations. In each case, strong redshift evolution is required. For example, the best-fit duty cycle in the energy-regulated model (blue) evolves from $\fduty \sim 0.8$ to $\fduty \sim 0.2$ (for $M_h \simeq 10^{10} \ \Msun$) over $4 \lesssim z \lesssim 8$ (middle panel), i.e., $\fduty \propto (1+z)^{-2}$. While a mass-dependent gradient $d\fduty/dM_h > 0$ is preferred at $1 \sigma$, our constraints are consistent with a mass \textit{independent} duty cycle at $2 \sigma$, as shown in Fig. \ref{fig:pdfs}. The dust-to-metal-ratio also must decline with redshift, by a factor of $\sim 2$ for $M_h \simeq 10^{10} \ \Msun$ halos over $4 \lesssim z \lesssim 8$ (right panel). Again, there is not  significant $\gtrsim 2\sigma$ evidence that a mass-dependent $\fdtmr$ is needed in the energy-regulated case. However, the shallower $f_{\ast} \propto M_h^{1/3}$ SFE in the momentum-regulated models (cyan) results in a more significant mass-dependence in $\fduty$. The inferred redshift evolution is comparable, though slightly weaker, in the momentum-regulated case than the energy-regulated case\footnote{Note that the absolute normalization of $f_{\ast}$, $\fduty$, and $\fdtmr$ are subject to revision given their dependence on parameters we hold fixed, e.g., the stellar metallicity and dust opacity, hence our focus only on the inferred mass and redshift scalings.}.

Figure \ref{fig:pdfs} further examines the statistical significance of the inferred mass and redshift-dependences of $\fduty$ and $\fdtmr$. Here, the posterior distribution of the power-law with mass ($\alpha$) and redshift ($\gamma$) dependences are shown, with dotted lines in each dimension indicating the limit in which no mass or redshift dependence is required by the data. The first and third columns focus on the mass-dependence inferred for $\fduty$ and $\fdtmr$, respectively, while the second and fourth columns address the redshift evolution. The marginalized 1-D constraints (shown along diagonal) are consistent with $\alpha_{\mathrm{duty}} = 0$ and $\alpha_{\mathrm{dtmr}} = 0$ in the energy-regulated (blue) case but not the momentum regulated case (cyan). In contrast, redshift evolution in both $\fduty$ and $\fdtmr$ is strongly required by the data (see second panel of bottom row for joint PDF) in both \evol\ models. Both $\fduty$ and $\fdtmr$ are degenerate with the amount of log-normal scatter in the dust column density at fixed $M_h$, parameterized via $\sigmaN$. The bimodality caused by $\sigmaN$ that we infer is robust, as the posterior distributions shown in Figure \ref{fig:pdfs} are computed using the last $\sim 100,000$ elements of a $\sim 10^6$ element MCMC chain, and do not change when computed with different subsets of the full chain. Note that the \texttt{evol} model put forth in this work is more amenable to large values of the scatter in the dust surface density, $0.1 \lesssim \sigmaN \lesssim 0.2$, than the \texttt{univ} model, and thus may help explain the evolution of the LAE fraction at $3 \lesssim z \lesssim 6$ \citep{Mirocha2020}. Best-fitting parameter values and their uncertainties are listed in Table \ref{tab:parameters} for reference.

\begin{table}%\scriptsize
\begin{tabular}{ | l | l | l | l | l ||}
\hline
parameter & \univ & \evole & \evolm & prior \\
\hline
$\log_{10}f_{\ast,10}$ & $-1.26^{+0.06}_{-0.02}$ & $-0.78^{+0.06}_{-0.29}$ & $-0.70^{+0.04}_{-0.20}$ & (-3, 0) \\
$\log_{10} M_{p,\ast}$ & $11.16^{+0.17}_{-0.19}$ & $10.78^{+0.38}_{-0.02}$ & $11.15^{+0.54}_{-0.10}$ & (9, 13) \\
$\alpha_{\ast,\mathrm{lo}}$ & $0.80^{+0.10}_{-0.14}$ & 2/3 & 1/3 & (0, 1.5) \\
$\alpha_{\ast,\mathrm{hi}}$ & $-0.53^{+0.24}_{-0.02}$ & $-0.38^{+0.10}_{-0.09}$ & $-0.55^{+0.12}_{-0.11}$ & (-3, 0.3) \\
$\gamma_{\ast,10}$ & 0 & 1 & 1/2 & (-3, 3) \\
\hline
$R_{d,10} / \mathrm{kpc}$ & $1.12^{+0.1}_{-0.08}$ & $0.86^{+0.66}_{-0.01}$ & $0.69^{+0.78}_{-0.13}$ & (0.1, 10) \\
$\log_{10} M_{p,d}$ & $12.01^{+0.31}_{-1.15}$ & $10.97^{+0.016}_{-0.49}$ & $11.22^{+0.23}_{-0.66}$ & (9, 13) \\
$\alpha_{d,\mathrm{lo}}$ & $0.69^{+0.16}_{-0.02}$ & $0.90^{+0.18}_{-0.02}$ & $1.03^{+0.04}_{-0.16}$ & (-2, 2) \\
$\alpha_{d,\mathrm{hi}}$ & $0.09^{+0.22}_{-0.07}$ & 1/3 & 1/3 & (-2, 2) \\
$\sigma_{\log_{10} N_d}$ & $\leq 0.08$ & $\leq 0.22$ & $\leq 0.20$ & (0, 1) \\
\hline
\hline
$\fdutynorm$  & 1   & $0.58^{+0.35}_{-0.042}$ &  $0.53^{+0.30}_{-0.01}$ & (0, 1) \\
$\fdutyalpha$ & 0   & $0.13^{+0.13}_{-0.090}$ & $0.28^{+0.21}_{-0.06}$ & (-2, 2) \\
$\fdutygamma$ & 0   & $-1.87^{+0.57}_{-0.30}$ & $-1.45^{+0.11}_{-0.83}$ & (-5, 5) \\
\hline
$\fdtmrnorm$  & 0.4 & $0.43^{+0.39}_{-0.02}$ & $0.29^{+0.57}_{-0.12}$ & (0, 1) \\
$\fdtmralpha$ & 0   & $0.22^{+0.02}_{-0.18}$ & $0.25^{+0.02}_{-0.20}$ & (-2, 2) \\
$\fdtmrgamma$ & 0   & $-1.32^{+0.43}_{-0.14}$ & $-1.34^{+0.39}_{-0.03}$ & (-5, 5) \\
\hline
\end{tabular}
\caption{{\bf Marginalized 68\% confidence intervals on the parameters of each model.} Values fixed in the fit are those without error-bars. Note that we only obtain upper limits on $\sigma_{\log_{10} N_d}$, which are listed at 99\% confidence.}
\label{tab:parameters}
\end{table}

\begin{figure*}
\begin{center}
\includegraphics[width=0.98\textwidth]{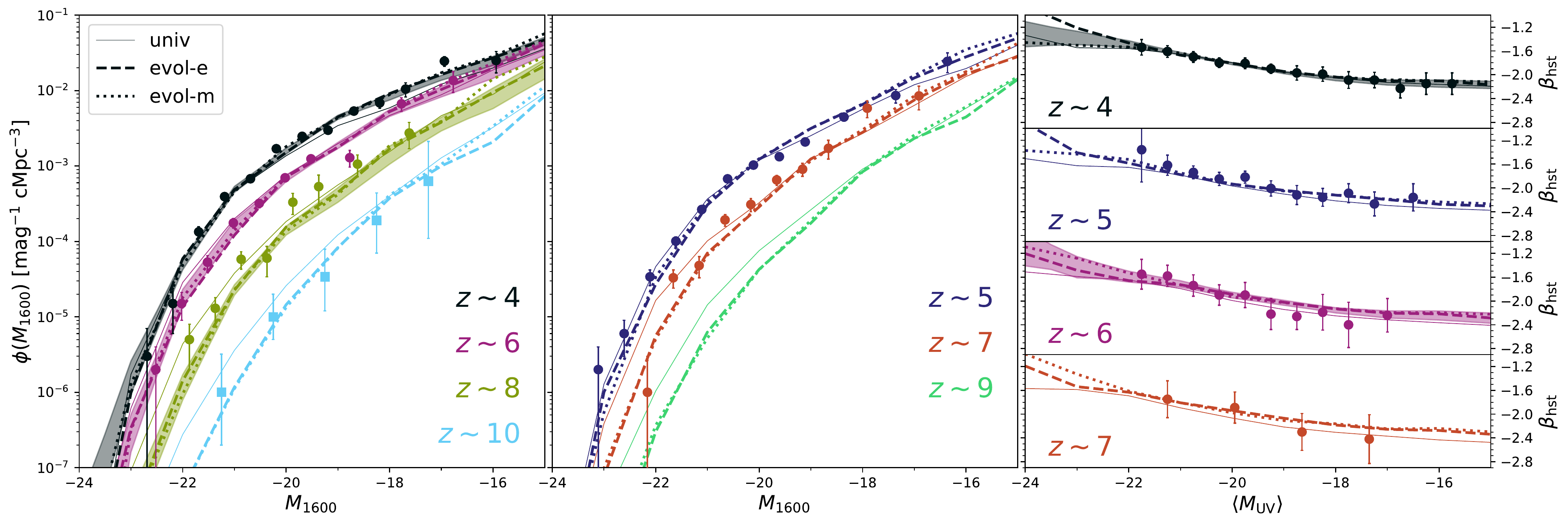}
\includegraphics[width=0.98\textwidth]{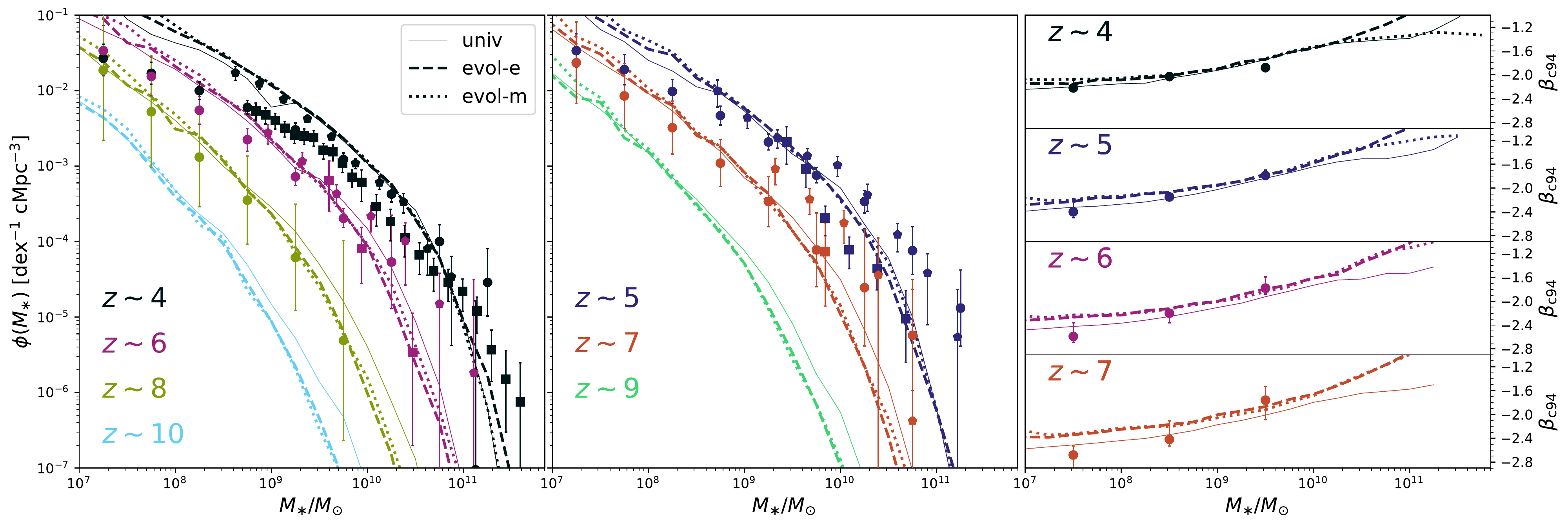}
\caption{{\bf Predictions for UVLFs and $\MUV$-$\beta$ relations are nearly indistinguishable between \texttt{univ} and \texttt{evol} models, in part by construction.} \textit{Top:} Left and center panels show UVLFs at a series of redshifts, while right-most panels show the evolution in the $\MUV$-$\beta$ relation over $4 \lesssim z \lesssim 7$, where $\beta$ is computed via mock HST photometry. Filled contours are included at redshifts included in the calibration, and indicate 68\% confidence reconstructions. \textit{Bottom:} Left and center panels show stellar mass functions at a series of redshifts, while right-most panels show the evolution in the $M_{\ast}$-$\beta$ relation over $4 \lesssim z \lesssim 7$, where $\beta$ is computed in the \citet{Calzetti1994} windows rather than mock HST photometry. Measurements included are as follows: UVLFs at $4 \lesssim z \lesssim 8$ from \citet{Bouwens2015} and $z\sim 10$ from \citet{Oesch2018}, $\MUV$-$\beta_{\mathrm{hst}}$ from \citet{Bouwens2014}, SMFs from \citet{Song2016} (circles), \citet{Stefanon2017} (squares), and \citet{Duncan2014} (pentagons), and $M_{\ast}$-$\beta_{\mathrm{c94}}$ relations from \citet{Finkelstein2012}.}
\label{fig:calib}
\end{center}
\end{figure*}

Despite the qualitatively different inputs to each model, they of course largely agree in their predictions for the UVLF and $\MUV$-$\beta$ relation, at least over the range of UV magnitudes and redshifts used in the calibration. In Figure \ref{fig:calib}, we examine this statement in more detail, showing the recovered UVLF and $\MUV$-$\beta$ relations (top), as well as predictions for the SMF and $M_{\ast}$-$\beta$ relations (bottom) over a broad range in brightness, mass, and redshift. The best-fit models do differ to some extent, but are consistent within the uncertainties of current measurements used in the calibration.

\begin{figure}
\begin{center}
\includegraphics[width=0.49\textwidth]{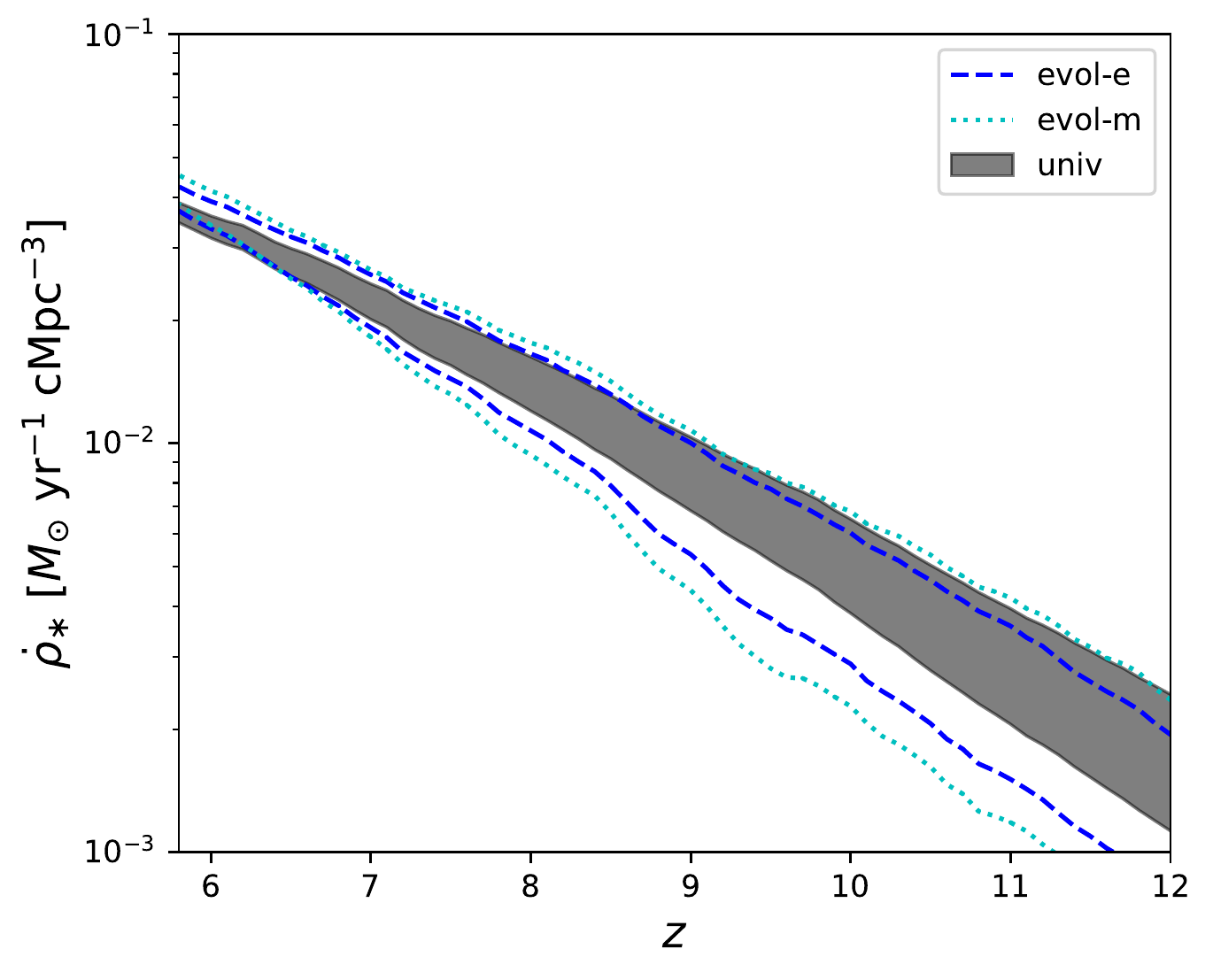}
\caption{{\bf Models in which $f_{\ast}$, $\fduty$, and $\fdtmr$ evolve, as in our \texttt{evol} models, are indistinguishable from \univ\ models in the cosmic SFRD, and thus cannot be distinguished via constraints from the mean reionization history or supernovae / gamma ray burst detection rates.} Shown above are the 68\%  reconstructions for the \univ\ (gray filled contours), energy-regulated (\evole; blue), and momentum-regulated (\evolm; cyan) models.}
\label{fig:sfrd}
\end{center}
\end{figure}

Predictions for the cosmic star formation rate density (SFRD) are also indistinguishable, as shown in Figure \ref{fig:sfrd}. As a result, independent probes of the SFRD via, e.g., the mean reionization history, will not obviously help distinguish these scenarios. However, this may only be true for models that neglect the inhomogeneous nature of reionization. For example, in extreme models with $\fduty \lesssim 0.1$, \citet{Hartley2016} find that bursty star formation can significantly boost the Thomson scattering optical depth of the cosmic microwave background. We find that $\fduty < 0.1$ is only realized for halos near the atomic cooling threshold (or below) at $z \gtrsim 8$.

\begin{figure*}
\begin{center}
\includegraphics[width=0.98\textwidth]{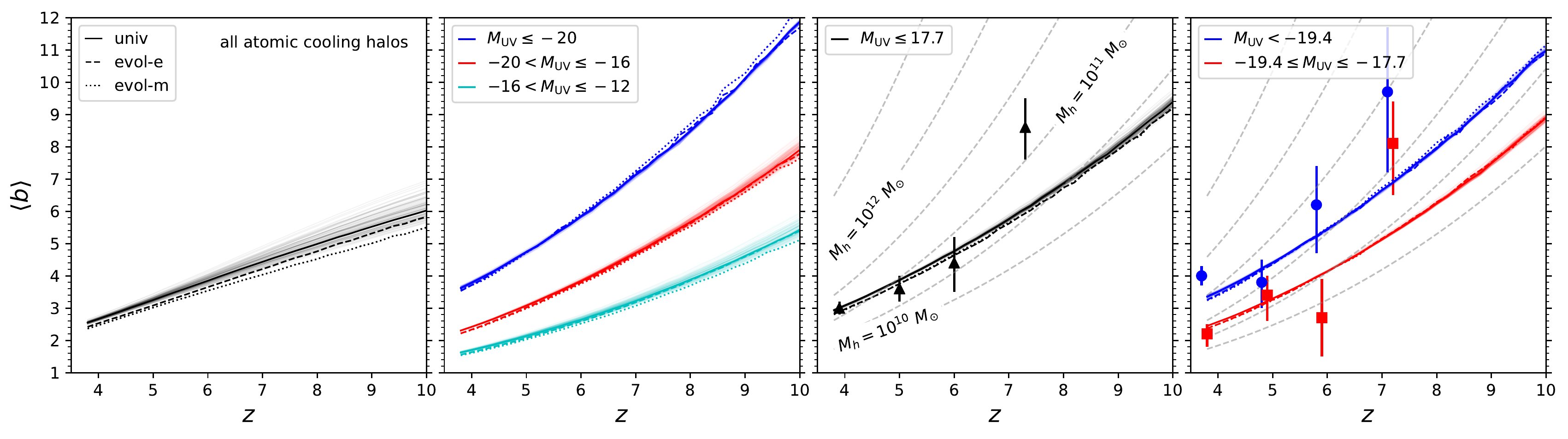}
\caption{{\bf Models in which $f_{\ast}$, $\fduty$, and $\fdtmr$ evolve, as in our \texttt{evol} models, predict weaker clustering of galaxies at high redshift than ``universal'' models.} \textit{Left:} Mean bias of all halos above the atomic cooling threshold for \texttt{univ} (solid) and \texttt{evol} (dashed, dotted) models as a function off redshift. \textit{Center Left:} Mean bias of galaxies in different coarse magnitude bins, including $\MUV \leq -20$ (blue), $-20 < \MUV \leq -16$ (red), and $-16 < \MUV \leq -12$ (cyan). \textit{Center Right:} Mean bias of galaxes with $\MUV < -17.7$, as in \citet{BaroneNugent2014}, with dashed lines indicating the bias of DM halos of fixed mass (every 0.5 dex). \textit{Right:} Mean bias split into the `bright` (blue) and `faint` (red) $\MUV$ bins used in \citet{BaroneNugent2014}. In each panel, we show 100 random samples of the posterior for the \univ\ model in semi-transparent lines, in order to indicate the level of uncertainty in the model calibration.}
\label{fig:bias}
\end{center}
\end{figure*}

Finally, given that the \texttt{univ} and \texttt{evol} models largely differ in the mass-to-light ratio of galaxies, we explore the degree to which clustering measurements can differentiate them. In Figure \ref{fig:bias}, we show the mean luminosity-weighted linear bias of galaxies in each model, defined as
\begin{equation}
	\langle b \rangle = \frac{\int_{\mmin}^{\mmax} dM_h \frac{dn}{dM_h} b(M_h) L(M_h) }{\int_{\mmin}^{\mmax} dM_h \frac{dn}{dM_h} L(M_h)}
\end{equation}
where $L$ is the $1600 \angstrom$ luminosity of galaxies in halos of mass $M_h$, and $b$ is the linear bias of a halo of mass $M_h$, which we compute using the fitting function provided by \citet{Tinker2010}.

We first average over all halos above the atomic cooling threshold (left), and then broken down into various coarse magnitude bins (left-center). The bias does vary among our models, though the difference is small, $\langle b_{\mathrm{univ}} \rangle - \langle b_{\mathrm{evol}} \rangle \lesssim 0.3$, at all redshifts when averaging over all atomic cooling halos, and even smaller for the different magnitude cuts. The difference between models is, as expected, greatest for the faintest galaxies at the highest redshifts, e.g., the $-16 < \MUV \leq -12$ bin (cyan lines in center-left panel). For galaxies brighter than $\MUV \lesssim -16$, best-fit model predictions differ in their bias predictions by only $\simeq 0.1$, which is comparable to the uncertainty in the predictions (semi-transpent lines show 100 random draws from the \univ\ model posterior disribution). As a result, intensity mapping measurements are likely the most promising approach to distinguishing models, as they capture all photons and thus compare most closely to our predictions for all atomic cooling halos, where models differ most.

We also compare our predictions to the constraints from \citet{BaroneNugent2014} (right two panels; Fig. \ref{fig:bias}). With a magnitude cut $\MUV \leq -17.7$ (center right; Fig. \ref{fig:bias}), all models agree well with the data at $4 \lesssim z \lesssim 6$. However, at $z \sim 7.2$, our models predict much weaker clustering than that reported by \citet{BaroneNugent2014}, which is true of other models as well \citep[e.g.,][]{Park2017}. We only find bias values $b \sim 8$ for $\MUV \lesssim -20$ objects (see blue curves, left-center panel), in agreement with current constraints on the clustering of very bright galaxies \citep{Hatfield2018}. This same general trend holds when dividing into the same `bright' and `faint' bins as \citet{BaroneNugent2014} (right panel; Fig. \ref{fig:bias}). In all cases, the difference in the predictions of \univ\ and \evol\ models is small, $\lesssim 0.1$, which is much smaller than the 1$\sigma$ uncertainties of current constraints \citep{BaroneNugent2014}.

%%
% DISCUSSION
%%
\section{Discussion} \label{sec:discussion}
In \S\ref{sec:results}, we showed that a suite of rest-ultraviolet observables can be matched both by ``universal'' SFE models, in which $f_{\ast}$ is a function of halo mass only, and energy- and momentum-regulated feedback models, in which case $f_{\ast} \propto M_h^{2/3} (1+z)$ and $f_{\ast} \propto M_h^{1/3} (1+z)^{1/2}$, respectively. However, in the latter two cases, strong evolution in the duty cycle and dust contents of galaxies are required to reconcile the model with observations. Though both varieties of models are in common use, no study to our knowledge has considered both classes of models in the same framework, allowed mass or redshift dependent duty cycles \textit{and} dust-to-metal ratios, or explored the extent to which these various cases can be diffentiated. In this section, we briefly discuss the implications of our work for stellar feedback models (\S\ref{sec:implications_feedback}), and the prospects for obtaining independent constraints on $\fduty$ (\S\ref{sec:evo_duty}), $\fdtmr$ (\S\ref{sec:evo_dtmr}), and galaxy clustering at high-$z$ (\S\ref{sec:clustering}).

%%
% What do we expect?
\subsection{Implications for stellar feedback models} \label{sec:implications_feedback}
The universality of the efficiency of star formation is a simple, but fundamental question in galaxy formation theory. Models with a universal SFE agree well with measurements at high redshift \citep{Sun2016,Mirocha2017,Tacchella2018}, and perhaps even over the entirety of cosmic history \citep{Trenti2010,Tacchella2013,Mason2015,Behroozi2013,Behroozi2019}. However, evolution in the SFE is a prediction of simple feedback models, in which star formation proceeds at a rate limited by the energy or momentum injection from supernovae explosions, which unbinds or expels accumulating gas that could otherwise form stars. Such arguments now underly many SAMs \citep[e.g.,][]{Somerville2012,Dayal2013,Furlanetto2017,Hutter2020}. Observationally, the situation is not so clear: some studies do support evolution in the closely-related stellar-mass-halo-mass relation, at least between $z \sim 2$ and $4$ \citep[e.g.,][]{Finkelstein2015b}, while clusering measurements support the universality of the SFE \citep{Harikane2018}, at least at $4 \lesssim z \lesssim 6$. As a result, it seems worthwhile to explore the extent to which other components and assumptions of semi-empirical models could hide evolution in the SFE, or alternatively, if the analytic feedback arguments underlying SAMs are missing some key ingredient(s) relevant to galaxy growth at high redshift.

In this work, we adopted models for the SFE representative of energy- and momentum-regulated feedback, as described in \citep{Furlanetto2017}. There are many variants of this argument in the literature \citep{Murray2005,Dayal2013,FaucherGiguere2013,Thompson2016,Hayward2017,Semenov2018,Krumholz2018}, though the rough mass and redshift scalings of the ``minimalist'' model of \citet{Furlanetto2017} generally emerge regardless of the details of the implementation (Furlanetto, submitted). As a result, the need for additional flexibility in the model, whether it be through $\fduty$ and $\fdtmr$ (as in this work), evolution in the dust optical depth, or otherwise, seems unlikely to disappear.

Though many different lines of reasoning result in similar predicted scalings between $f_{\ast}$, $M_h$, and $z$, it is not unreasonable to expect these predictions to fail to match measurements in detail. For example, all are based on smooth, inflow-driven star formation, and that a single mode of feedback regulates the star formation in all galaxies, at all times with the same coupling efficiency (i.e., fraction of supernova energy or momentum that couples to the interstellar medium). These assumptions are clearly idealized. For example, both modes of feedback may operate in rapid succession, as early episodes of star formation clear out channels for subsequent episodes, and thus minimize radiative losses. This of course has implications for the treatment of $\fduty$ in our model, which we assumed to be independent of the mode of feedback. Despite this potential for additional complexity, it is not obvious how relaxing  the assumptions of the model or explicitly linking feedback to burstiness would eliminate the predicted redshift dependence of the SFE.

A natural question to ask in this context is: \textit{how well do we expect analytic feedback models to work}? Given that \univ\ and \evol\ models differ more than the \evole\ and \evolm\ models differ from each other, it would thus be interesting to determine under what conditions one can archive a universal SFE from within a feedback-regulated framework, if burstiness can be incorporated self-consistently, and if so, how it evolves with redshift and/or halo mass.

%%
% fduty < 1?
\subsection{Implications of $\fduty$ evolution} \label{sec:evo_duty}
Naively, one can combat ``over-luminous'' galaxies at high-$z$ by assuming that each galaxy is only actively star-forming some fraction $\fduty < 1$ of the time \citep[see, e.g.,][]{Trenti2010,Wyithe2014}. In our framework, this route poses a challenge to self-consistency, since one of the key assumptions underlying the energy-regulated model is continuous star formation. As a result, our implementation of the energy-regulated model with $\fduty < 1$ effectively assumes that feedback operates as if star formation were continuous, despite star formation occurring only sporadically in any individual galaxy.

This sounds problematic, but another assumption may offset this apparent self-contradiction: the simplest analytic feedback arguments also assume that feedback acts instantaneously, i.e., that the energy injection rate from supernovae $\dot{E}_{\mathrm{SNe}}(t) \propto \mathrm{SFR}(t)$. In reality, feedback is delayed relative to star formation -- the energy and/or momentum injection rate from supernovae explosions will reflect the SFR of galaxy $\sim 5-30$ Myr earlier \citep[e.g.,][]{Mutch2016,Orr2018}. Provided that the duty cycle is not so low that galaxies regularly go many tens of Myr between star-forming episodes, it may not be entirely unreasonable for feedback to remain effective even without continuous star formation. A proper semi-analytic modeling treatment would include the delay-time distribution of supernovae explicitly, though it is not clear how to meaningfully incoporate this effect into analytic models like those of \citet{Furlanetto2017}, nor is it obvious that doing so would change the expected scalings of $f_{\ast}$ with halo mass and time.

Self-consistency issues aside, the behaviour of $\fduty$ inferred by the \texttt{evol} models has some desireable features. There is some empirical evidence for bursty star formation, e.g., variability on short $\sim 10-100$ Myr timescales affects the ratio of H$\alpha$ to non-ionizing UV continuum emission \citep{Weisz2012,Broussard2019,Emami2019}. Recent measurements targeting relatively bright $z\sim 4-5$ galaxies suggest fairly smooth recent star formation histories \citep{Smit2016}, which is not inconsistent with the \evol\ model predictions given that our best-fit relations converge toward smoother star formation at lower redshift and higher mass. Future efforts with \textit{JWST} will help push these constraints to higher redshift and fainter galaxy populations, where our models predict more burstiness. Qualitatively, the burstiness trends we infer are expected, since dynamical times in high-$z$ galaxies can be shorter than the timescale over which supernovae feedback is injected into the interstellar medium \citep{FaucherGiguere2018}. Numerical simulations of galaxy formation do exhibit highly variable SFHs \citep[see, e.g.,][]{Hopkins2018}, the details of which vary noticeably among models \citep{Iyer2020}. Our approach is likely overly-simplistic, given that we randomly toggle the SFR of galaxies on and off. We plan to explore more complex variability in future work.

%%
% Talk about Popping et al. models, ALMA observations, AGB v. SNe
\subsection{Implications of $\fdtmr$ evolution} \label{sec:evo_dtmr}
At first glance it might seem like the increasingly efficient star formation of \evol\ models could naturally avoid making over-luminous galaxies (with respect to UVLFs), given that star formation yields metals, and thus more potential for dust production. However, we find that the brightening associated with an increasing SFE outpaces any dimming from additional dust production (colours are slightly redder, but additional dimming is insignificant). As a result, we confirm the need for a decline in dust production or retention with redshift pointed out by many other authors in recent years \citep[e.g.,][]{Somerville2012,Yung2019a,Vogelsberger2019,Qiu2019}. Whereas most previous work has invoked purely redshift-dependent modifications to the model, we also allowed the possibility of a halo mass dependence. However, we find that the evidence for redshift evolution in $\fdtmr$ (and $\fduty$) is much stronger than the evidence for a halo mass dependence (see Fig. \ref{fig:pdfs}).

Just as there are physically-motivated reasons to expect a decline in $\fduty$ with redshift (see \S\ref{sec:evo_duty}), there are many reasons to expect $\fdtmr$ to decline with redshift. The Universe is too young at $z \gtrsim 6$ for the most efficient dust producers -- asymptotic giant branch (AGB) stars -- to have emerged, leaving supernovae as the only plausible site of dust production. A purely redshift-dependent $\fdtmr$ could indicate that, no matter when a halo began forming stars, AGB stars dictate the dust contents of galaxies as soon as they appear on the scene. The mild $M_h$-dependence in $\fdtmr$ we infer could simply indicate that the most massive galaxies have been forming stars longer, and thus will have more AGB stars at late times than galaxies hosted by smaller halos, or an additional dependence on metallicity, which we have not attempted to model explicitly \citep[see, e.g.,][]{Li2019}.

%%
% CLUSTERING
\subsection{Prospects for clustering constraints} \label{sec:clustering}
The \univ\ and \evol\ model predictions differ by $\langle b_{\mathrm{univ}} \rangle - \langle b_{\mathrm{evol}} \rangle \simeq 0.1-0.3$, depending on redshift and magnitude cut. This corresponds to a difference in predictions for the masses of halos hosting high-$z$ galaxies of only $\lesssim 0.1$ dex. The uncertainties in current measurements are at least as large, approaching $\sigma_b \simeq 1$, at $z \gtrsim 6$ \citep{BaroneNugent2014}, and are thus unable to distinguish the \univ\ and \evol\ models. Though a direct comparison with a broader array of constraints is difficult, as we present results in terms of the bias rather than angular correlation function or halo masses, typical uncertainties for halo masses in recent years are $\sim 0.2-0.5$ dex \citep[see, e.g.,][]{Harikane2016,Hatfield2018,Khostovan2019}, and much tighter in some cases \citep{Harikane2018}. However, most of the aforementioned constraints are for bright galaxies living in massive $\sim 10^{12} \ M_{\odot}$ halos, which likely do not abide by the rules relevant to low mass galaxies, below the peak of the SFE $\Mpeak \sim \mathrm{few} \times 10^{11} \ M_{\odot}$. As a result, we caution against interpreting our bias predictions for the most massive halos, which carry the SFE redshift evolution expected in stellar feedback arguments despite a different mass-dependence. Future measurements with \textit{JWST} are particularly appealing in this respect, as they are expected to provide high significance detections of galaxy clustering out to $z \sim 10$, for objects as faint as $\MUV \sim -18$ \citep{Endsley2020}, which corresponds to $M_h \sim 10^{10} \ M_{\odot}$ in most empirical models. Future pure parallel programs could also be very complementary to angular correlation function techniques in this context \citep{Robertson2010}.

% CAVEATS
\subsection{Caveats}
The simplicity of our models could be a source of additional uncertainty or bias in our inferred $f_{\ast}$, $\fdtmr$, and $\fduty$ relationships. For example, we build idealized halo growth trajectories, essentially by neglecting mergers. A model built on merger trees would have more diversity in star formation histories (and thus dust production) than ours, which imposes log-normal scatter in SFRs (at fixed $M_h$) by-hand. Some of the behavior in $\fduty$ and $\fdtmr$ could thus emerge naturally given the diversity in halo assembly alone. Along these lines, we have imposed redshift evolution in $f_{\ast}$ appropriate for stellar feedback models at \textit{all} masses, even those above the peak of the SFE curve. This may not be appropriate given the clear change in relevant physical processes implied by the departure from a pure power-law, and will certainly affect the inferred behaviour of $\fstar$, $\fduty$, and $\fdtmr$ at $M_h \gtrsim 5 \times 10^{11} \ M_{\odot}$.

It is also possible that the inclusion of late-time or alternative constraints in our model calibration could affect our conclusions. For example, we have opted to leave SMF measurements out of the model calibration, given the limited long-wavelength coverage of current SMF estimates. There is a clear tension here, as our models all predict steep SMFs, in line with the \citet{Duncan2014} measurements but in tension with those of \citet{Song2016} and \citet{Stefanon2017}, while agreeing well with most UVLF constraints. Future SMF measurements with \textit{JWST} will thus provide an important alternative to rest-ultraviolet inference procedures like ours, that may qualitatively shift the inferred behaviour of key model inputs -- especially $\fdtmr$.

Lastly, our exploration of model parameter space in this paper is by no means exhaustive. There are likely other ways to accommodate the redshift evolution introduced by feedback-regulated models, e.g., appeals to mass and time-dependent changes in dust composition and/or geometry, both of which are expected to some extent \citep[see, e.g.,][]{Popping2017,Narayanan2018}. However, the two common cases we have explored serve to illustrate the fundamental challenge of distinguishing different galaxy formation scenarios at high redshift. One could arrive at similar conclusions by flexibly parameterizing $f_{\ast}$, $\fduty$, and $\fdtmr$, and noting the broadening of posterior distributions in a figure analogous to our Figure \ref{fig:pdfs}. Instead, we have adopted three specific cases representative of models in the recent literature, to better illustrate the physical meaning of differrent points in this high-dimensional parameter space.

%%
% CONCLUSIONS
%%
\section{Conclusions} \label{sec:conclusions}
While models that adopt a universal star formation efficiency can fit current constraints on the high-$z$ galaxy population well, simple feedback arguments predict that the SFE should grow with increasing redshift. Even though the goal of universal SFE models is not to be explanatory -- in fact, part of the allure is to be able to make new predictions without a physical model for galaxies -- one would ideally be able to distinguish empirically- and physically-motivated models. In this work, we first generalized a common feedback-regulated model for star formation to allow evolution also in the duty cycle and dust production efficiency (our \evol\ models), and then compared its predictions to those derived from a semi-empirical framework with a time-independent star formation efficiency (our \univ\ model). Our conclusions can be summarized as follows:
\begin{itemize}
	\item Imposing redshift evolution in the efficiency of star formation at a level predicted by common feedback models, $f_{\ast} \propto M_h^{1/3-2/3} (1+z)^{1/2-1}$, results in overly-luminous high-$z$ galaxies, whose colours become \textit{redder} with $z$ (at fixed $\MUV$) due to the corresponding boost in dust production, in tension with current constraints. Furthermore, if dust scale lengths are related to halo virial radii, reddening becomes even more extreme, resulting in steep $\MUV$-$\beta$ relationships and a dearth of UV-bright galaxies (see Figures \ref{fig:zevol_sfe}-\ref{fig:zevol_Rd}).
	\item To counter these effects, we allow the star formation duty cycle, $\fduty$, and dust-to-metal ratio, $\fdtmr$, to vary freely in our semi-empirical modeling, while holding the behavior of the SFE fixed with values appropriate for energy- and momentum-regulated feedback. We find that $\fduty$ and $\fdtmr$ must both decline rapidly with $z$ in order to reconcile the feedback-regulated models with UVLFs and UV colours at $4 \lesssim z \lesssim 8$, roughly as $\propto (1+z)^{-3/2}$, i.e., on a Hubble timescale (see Figures \ref{fig:recon}, \ref{fig:pdfs}, and Table \ref{tab:parameters}). An additional dependence on $M_h$ is needed in $\fduty$ for momentum-regulated models, but only preferred at the $\sim 1\sigma$ level for $\fdtmr$ and in each quantity for energy-regulated models. Evolution in $\fdtmr$ mitigates the over-reddening problem caused by rapid size evolution, $R_d \propto \rvir$, while $\fduty$ evolution reduces the typical luminosity of galaxies at fixed abundance, counteracting the growing efficiency of star formation in energy-regulated models.
	\item By construction, the feedback-regulated models with evolving $\fduty$ and $\fdtmr$ are nearly indistinguishable from ``universal'' models, in which neither the SFE or dust properties of galaxies evolve with time (see Fig. \ref{fig:calib}). Intensity mapping experiments may provide an important discriminant among models, as the mean galaxy bias differs most when averaging over the entire galaxy population (e.g., $\langle b_{\mathrm{univ}} \rangle - \langle b_{\mathrm{evol}} \rangle \simeq 0.1-0.3$ at $z \sim 4-10$; see Figure \ref{fig:bias}). Independent constraints on the core inputs of the model, i.e., the duty cyle and dust-to-metal ratio (or dust-mass-stellar-mass relation), are the only other obvious way to distinguish the \univ\ and \evol\ models explored in this work.
	\item Our approach effectively assumes that feedback operates as if star formation were occurring continuously, though the standard energy-regulated feedback argument assumes smooth inflow-driven star formation. Provided that $\fduty \simeq 1$, supernovae may still be able to sustain feedback during the brief lulls in a galaxy's star formation history, given the delay between star formation and supernovae explosions. Our models require $\fduty \sim 0.2$ in $M_h \sim 10^{10} \ \Msun$ halos at $z \sim 8$, and thus may strain such arguments. Moving forward, it would be useful to explore the extent to which burstiness can be incorporated in analytic frameworks, and perhaps compare to $\fduty$ estimates from \textit{ab initio} galaxy formation simulations, in which burstiness arises naturally.
\end{itemize}

J.M. acknowledges helpful feedback on an early draft from Steve Furlanetto and Louis Abramson, stimulating conversations with Chris Matzner and Norm Murray, support through a CITA National Fellowship, and the anonymous referee for feedback that helped improve the paper. Computations were made on the supercomputer Cedar at Simon Fraser University managed by Compute Canada. The operation of this supercomputer is funded by the Canada Foundation for Innovation (CFI).

\textit{Software:} numpy \citep{numpy}, scipy \citep{scipy}, matplotlib \citep{matplotlib}, h5py\footnote{\url{http://www.h5py.org/}}, and mpi4py \citep{mpi4py1}.

\textit{Data Availability:} The data underlying this article is available upon request, but can also be re-generated from scratch using the publicly available \textsc{ares} code.

\bibliography{references}

\newcommand{\noop}[1]{}
\begin{thebibliography}{76}
\expandafter\ifx\csname natexlab\endcsname\relax\def\natexlab#1{#1}\fi

\bibitem[{Barone-Nugent {et~al}\mbox{.}(2014)Barone-Nugent, Trenti, Wyithe,
  Bouwens, Oesch, Illingworth, Carollo, Su, Stiavelli, Labb{\'e}, \&
  Van~Dokkum}]{BaroneNugent2014}
Barone-Nugent R.~L. {et~al.}, 2014, \apj, 793, 17

\bibitem[{{Behroozi} {et~al}\mbox{.}(2019){Behroozi}, {Wechsler}, {Hearin}, \&
  {Conroy}}]{Behroozi2019}
{Behroozi} P. {et~al.}, 2019, \mnras, 488, 3143

\bibitem[{Behroozi {et~al}\mbox{.}(2013)Behroozi, Wechsler, \&
  Conroy}]{Behroozi2013}
Behroozi P.~S., Wechsler R.~H., Conroy C., 2013, \apj, 770, 57

\bibitem[{{Bouwens} {et~al}\mbox{.}(2011){Bouwens}, {Illingworth}, {Oesch},
  {Labb{\'e}}, {Trenti}, {van Dokkum}, {Franx}, {Stiavelli}, {Carollo},
  {Magee}, \& {Gonzalez}}]{Bouwens2011}
{Bouwens} R.~J. {et~al.}, 2011, \apj, 737, 90

\bibitem[{{Bouwens} {et~al}\mbox{.}(2014){Bouwens}, {Illingworth}, {Oesch},
  {Labb{\'e}}, {van Dokkum}, {Trenti}, {Franx}, {Smit}, {Gonzalez}, \&
  {Magee}}]{Bouwens2014}
{Bouwens} R.~J. {et~al.}, 2014, \apj, 793, 115

\bibitem[{{Bouwens} {et~al}\mbox{.}(2015){Bouwens}, {Illingworth}, {Oesch},
  {Trenti}, {Labb{\'e}}, {Bradley}, {Carollo}, {van Dokkum}, {Gonzalez},
  {Holwerda}, {Franx}, {Spitler}, {Smit}, \& {Magee}}]{Bouwens2015}
{Bouwens} R.~J. {et~al.}, 2015, \apj, 803, 34

\bibitem[{{Bowler} {et~al}\mbox{.}(2020){Bowler}, {Jarvis}, {Dunlop}, {McLure},
  {McLeod}, {Adams}, {Milvang-Jensen}, \& {McCracken}}]{Bowler2020}
{Bowler} R.~A.~A. {et~al.}, 2020, \mnras, 493, 2059

\bibitem[{{Broussard} {et~al}\mbox{.}(2019){Broussard}, {Gawiser}, {Iyer},
  {Kurczynski}, {Somerville}, {Dav{\'e}}, {Finkelstein}, {Jung}, \&
  {Pacifici}}]{Broussard2019}
{Broussard} A. {et~al.}, 2019, \apj, 873, 74

\bibitem[{{Calzetti} {et~al}\mbox{.}(1994){Calzetti}, {Kinney}, \&
  {Storchi-Bergmann}}]{Calzetti1994}
{Calzetti} D., {Kinney} A.~L., {Storchi-Bergmann} T., 1994, \apj, 429, 582

\bibitem[{Dalcín {et~al}\mbox{.}(2005)Dalcín, Paz, \& Storti}]{mpi4py1}
Dalcín L., Paz R., Storti M., 2005, Journal of Parallel and Distributed
  Computing, 65, 1108

\bibitem[{Dayal {et~al}\mbox{.}(2013)Dayal, Dunlop, Maio, \&
  Ciardi}]{Dayal2013}
Dayal P. {et~al.}, 2013, \mnras, 434, 1486

\bibitem[{{Duncan} {et~al}\mbox{.}(2014){Duncan}, {Conselice}, {Mortlock},
  {Hartley}, {Guo}, {Ferguson}, {Dav{\'e}}, {Lu}, {Ownsworth}, {Ashby},
  {Dekel}, {Dickinson}, {Faber}, {Giavalisco}, {Grogin}, {Kocevski},
  {Koekemoer}, {Somerville}, \& {White}}]{Duncan2014}
{Duncan} K. {et~al.}, 2014, \mnras, 444, 2960

\bibitem[{{Eldridge} \& {Stanway}(2009)}]{Eldridge2009}
{Eldridge} J.~J., {Stanway} E.~R., 2009, \mnras, 400, 1019

\bibitem[{{Emami} {et~al}\mbox{.}(2019){Emami}, {Siana}, {Weisz}, {Johnson},
  {Ma}, \& {El-Badry}}]{Emami2019}
{Emami} N. {et~al.}, 2019, \apj, 881, 71

\bibitem[{{Endsley} {et~al}\mbox{.}(2020){Endsley}, {Behroozi}, {Stark},
  {Williams}, {Robertson}, {Rieke}, {Gottl{\"o}ber}, \& {Yepes}}]{Endsley2020}
{Endsley} R. {et~al.}, 2020, \mnras, 493, 1178

\bibitem[{{Faucher-Gigu{\`e}re}(2018)}]{FaucherGiguere2018}
{Faucher-Gigu{\`e}re} C.-A., 2018, \mnras, 473, 3717

\bibitem[{{Faucher-Gigu{\`e}re} {et~al}\mbox{.}(2013){Faucher-Gigu{\`e}re},
  {Quataert}, \& {Hopkins}}]{FaucherGiguere2013}
{Faucher-Gigu{\`e}re} C.-A., {Quataert} E., {Hopkins} P.~F., 2013, \mnras, 433,
  1970

\bibitem[{{Finkelstein} {et~al}\mbox{.}(2012){Finkelstein}, {Papovich},
  {Salmon}, {Finlator}, {Dickinson}, {Ferguson}, {Giavalisco}, {Koekemoer},
  {Reddy}, {Bassett}, {Conselice}, {Dunlop}, {Faber}, {Grogin}, {Hathi},
  {Kocevski}, {Lai}, {Lee}, {McLure}, {Mobasher}, \&
  {Newman}}]{Finkelstein2012}
{Finkelstein} S.~L. {et~al.}, 2012, \apj, 756, 164

\bibitem[{{Finkelstein} {et~al}\mbox{.}(2015{\natexlab{a}}){Finkelstein},
  {Ryan}, {Papovich}, {Dickinson}, {Song}, {Somerville}, {Ferguson}, {Salmon},
  {Giavalisco}, {Koekemoer}, {Ashby}, {Behroozi}, {Castellano}, {Dunlop},
  {Faber}, {Fazio}, {Fontana}, {Grogin}, {Hathi}, {Jaacks}, {Kocevski},
  {Livermore}, {McLure}, {Merlin}, {Mobasher}, {Newman}, {Rafelski}, {Tilvi},
  \& {Willner}}]{Finkelstein2015}
{Finkelstein} S.~L. {et~al.}, 2015{\natexlab{a}}, \apj, 810, 71

\bibitem[{{Finkelstein} {et~al}\mbox{.}(2015{\natexlab{b}}){Finkelstein},
  {Song}, {Behroozi}, {Somerville}, {Papovich}, {Milosavljevi{\'c}}, {Dekel},
  {Narayanan}, {Ashby}, {Cooray}, {Fazio}, {Ferguson}, {Koekemoer}, {Salmon},
  \& {Willner}}]{Finkelstein2015b}
{Finkelstein} S.~L. {et~al.}, 2015{\natexlab{b}}, \apj, 814, 95

\bibitem[{{Foreman-Mackey} {et~al}\mbox{.}(2013){Foreman-Mackey}, {Hogg},
  {Lang}, \& {Goodman}}]{ForemanMackey2013}
{Foreman-Mackey} D. {et~al.}, 2013, \pasp, 125, 306

\bibitem[{{Furlanetto} {et~al}\mbox{.}(2017){Furlanetto}, {Mirocha}, {Mebane},
  \& {Sun}}]{Furlanetto2017}
{Furlanetto} S.~R. {et~al.}, 2017, \mnras, 472, 1576

\bibitem[{{Grogin} {et~al}\mbox{.}(2011){Grogin}, {Kocevski}, {Faber},
  {Ferguson}, {Koekemoer}, {Riess}, {Acquaviva}, {Alexander}, {Almaini},
  {Ashby}, {Barden}, {Bell}, {Bournaud}, {Brown}, {Caputi}, {Casertano},
  {Cassata}, {Castellano}, {Challis}, {Chary}, {Cheung}, {Cirasuolo},
  {Conselice}, {Roshan Cooray}, {Croton}, {Daddi}, {Dahlen}, {Dav{\'e}}, {de
  Mello}, {Dekel}, {Dickinson}, {Dolch}, {Donley}, {Dunlop}, {Dutton}, {Elbaz},
  {Fazio}, {Filippenko}, {Finkelstein}, {Fontana}, {Gardner}, {Garnavich},
  {Gawiser}, {Giavalisco}, {Grazian}, {Guo}, {Hathi}, {H{\"a}ussler},
  {Hopkins}, {Huang}, {Huang}, {Jha}, {Kartaltepe}, {Kirshner}, {Koo}, {Lai},
  {Lee}, {Li}, {Lotz}, {Lucas}, {Madau}, {McCarthy}, {McGrath}, {McIntosh},
  {McLure}, {Mobasher}, {Moustakas}, {Mozena}, {Nandra}, {Newman}, {Niemi},
  {Noeske}, {Papovich}, {Pentericci}, {Pope}, {Primack}, {Rajan},
  {Ravindranath}, {Reddy}, {Renzini}, {Rix}, {Robaina}, {Rodney}, {Rosario},
  {Rosati}, {Salimbeni}, {Scarlata}, {Siana}, {Simard}, {Smidt}, {Somerville},
  {Spinrad}, {Straughn}, {Strolger}, {Telford}, {Teplitz}, {Trump}, {van der
  Wel}, {Villforth}, {Wechsler}, {Weiner}, {Wiklind}, {Wild}, {Wilson},
  {Wuyts}, {Yan}, \& {Yun}}]{Grogin2011}
{Grogin} N.~A. {et~al.}, 2011, \apjs, 197, 35

\bibitem[{Harikane {et~al}\mbox{.}(2016)Harikane, Ouchi, Ono, More, Saito, Lin,
  Coupon, Shimasaku, Shibuya, Price, Lin, Hsieh, Ishigaki, Komiyama, Silverman,
  Takata, Tamazawa, \& Toshikawa}]{Harikane2016}
Harikane Y. {et~al.}, 2016, \apj, 821, 123

\bibitem[{{Harikane} {et~al}\mbox{.}(2018){Harikane}, {Ouchi}, {Ono}, {Saito},
  {Behroozi}, {More}, {Shimasaku}, {Toshikawa}, {Lin}, {Akiyama}, {Coupon},
  {Komiyama}, {Konno}, {Lin}, {Miyazaki}, {Nishizawa}, {Shibuya}, \&
  {Silverman}}]{Harikane2018}
{Harikane} Y. {et~al.}, 2018, \pasj, 70, S11

\bibitem[{{Hartley} \& {Ricotti}(2016)}]{Hartley2016}
{Hartley} B., {Ricotti} M., 2016, \mnras, 462, 1164

\bibitem[{{Hatfield} {et~al}\mbox{.}(2018){Hatfield}, {Bowler}, {Jarvis}, \&
  {Hale}}]{Hatfield2018}
{Hatfield} P.~W. {et~al.}, 2018, \mnras, 477, 3760

\bibitem[{Hayward \& Hopkins(2017)}]{Hayward2017}
Hayward C.~C., Hopkins P.~F., 2017, \mnras, 465, 1682

\bibitem[{{Hopkins} {et~al}\mbox{.}(2018){Hopkins}, {Wetzel}, {Kere{\v{s}}},
  {Faucher-Gigu{\`e}re}, {Quataert}, {Boylan-Kolchin}, {Murray}, {Hayward},
  {Garrison-Kimmel}, {Hummels}, {Feldmann}, {Torrey}, {Ma},
  {Angl{\'e}s-Alc{\'a}zar}, {Su}, {Orr}, {Schmitz}, {Escala}, {Sanderson},
  {Grudi{\'c}}, {Hafen}, {Kim}, {Fitts}, {Bullock}, {Wheeler}, {Chan},
  {Elbert}, \& {Narayanan}}]{Hopkins2018}
{Hopkins} P.~F. {et~al.}, 2018, \mnras, 480, 800

\bibitem[{Hunter(2007)}]{matplotlib}
Hunter J.~D., 2007, Computing in Science \& Engineering, 9, 90

\bibitem[{{Hutter} {et~al}\mbox{.}(2020){Hutter}, {Dayal}, {Yepes},
  {Gottl{\"o}ber}, {Legrand}, \& {Ucci}}]{Hutter2020}
{Hutter} A. {et~al.}, 2020, arXiv e-prints, arXiv:2004.08401

\bibitem[{{Illingworth} {et~al}\mbox{.}(2013){Illingworth}, {Magee}, {Oesch},
  {Bouwens}, {Labb{\'e}}, {Stiavelli}, {van Dokkum}, {Franx}, {Trenti},
  {Carollo}, \& {Gonzalez}}]{Illingworth2013}
{Illingworth} G.~D. {et~al.}, 2013, \apjs, 209, 6

\bibitem[{{Iyer} {et~al}\mbox{.}(2020){Iyer}, {Tacchella}, {Genel}, {Hayward},
  {Hernquist}, {Brooks}, {Caplar}, {Dav{\'e}}, {Diemer}, {Forbes}, {Gawiser},
  {Somerville}, \& {Starkenburg}}]{Iyer2020}
{Iyer} K.~G. {et~al.}, 2020, \mnras, 498, 430

\bibitem[{{Khostovan} {et~al}\mbox{.}(2019){Khostovan}, {Sobral}, {Mobasher},
  {Matthee}, {Cochrane}, {Chartab}, {Jafariyazani}, {Paulino-Afonso}, {Santos},
  \& {Calhau}}]{Khostovan2019}
{Khostovan} A.~A. {et~al.}, 2019, \mnras, 489, 555

\bibitem[{{Koekemoer} {et~al}\mbox{.}(2011){Koekemoer}, {Faber}, {Ferguson},
  {Grogin}, {Kocevski}, {Koo}, {Lai}, {Lotz}, {Lucas}, {McGrath}, {Ogaz},
  {Rajan}, {Riess}, {Rodney}, {Strolger}, {Casertano}, {Castellano}, {Dahlen},
  {Dickinson}, {Dolch}, {Fontana}, {Giavalisco}, {Grazian}, {Guo}, {Hathi},
  {Huang}, {van der Wel}, {Yan}, {Acquaviva}, {Alexander}, {Almaini}, {Ashby},
  {Barden}, {Bell}, {Bournaud}, {Brown}, {Caputi}, {Cassata}, {Challis},
  {Chary}, {Cheung}, {Cirasuolo}, {Conselice}, {Roshan Cooray}, {Croton},
  {Daddi}, {Dav{\'e}}, {de Mello}, {de Ravel}, {Dekel}, {Donley}, {Dunlop},
  {Dutton}, {Elbaz}, {Fazio}, {Filippenko}, {Finkelstein}, {Frazer}, {Gardner},
  {Garnavich}, {Gawiser}, {Gruetzbauch}, {Hartley}, {H{\"a}ussler},
  {Herrington}, {Hopkins}, {Huang}, {Jha}, {Johnson}, {Kartaltepe},
  {Khostovan}, {Kirshner}, {Lani}, {Lee}, {Li}, {Madau}, {McCarthy},
  {McIntosh}, {McLure}, {McPartland}, {Mobasher}, {Moreira}, {Mortlock},
  {Moustakas}, {Mozena}, {Nandra}, {Newman}, {Nielsen}, {Niemi}, {Noeske},
  {Papovich}, {Pentericci}, {Pope}, {Primack}, {Ravindranath}, {Reddy},
  {Renzini}, {Rix}, {Robaina}, {Rosario}, {Rosati}, {Salimbeni}, {Scarlata},
  {Siana}, {Simard}, {Smidt}, {Snyder}, {Somerville}, {Spinrad}, {Straughn},
  {Telford}, {Teplitz}, {Trump}, {Vargas}, {Villforth}, {Wagner}, {Wand ro},
  {Wechsler}, {Weiner}, {Wiklind}, {Wild}, {Wilson}, {Wuyts}, \&
  {Yun}}]{Koekemoer2011}
{Koekemoer} A.~M. {et~al.}, 2011, \apjs, 197, 36

\bibitem[{{Krumholz} {et~al}\mbox{.}(2018){Krumholz}, {Burkhart}, {Forbes}, \&
  {Crocker}}]{Krumholz2018}
{Krumholz} M.~R. {et~al.}, 2018, \mnras, 477, 2716

\bibitem[{{Li} {et~al}\mbox{.}(2019){Li}, {Narayanan}, \& {Dav{\'e}}}]{Li2019}
{Li} Q., {Narayanan} D., {Dav{\'e}} R., 2019, \mnras, 490, 1425

\bibitem[{Livermore {et~al}\mbox{.}(2018)Livermore, Trenti, Bradley, Bernard,
  Holwerda, Mason, \& Treu}]{Livermore2018}
Livermore R.~C. {et~al.}, 2018, \apjl, 861, L17

\bibitem[{{Mason} {et~al}\mbox{.}(2015){Mason}, {Trenti}, \&
  {Treu}}]{Mason2015}
{Mason} C.~A., {Trenti} M., {Treu} T., 2015, \apj, 813, 21

\bibitem[{McLeod {et~al}\mbox{.}(2016)McLeod, McLure, \& Dunlop}]{McLeod2016}
McLeod D.~J., McLure R.~J., Dunlop J.~S., 2016, \mnras, 459, 3812

\bibitem[{{Mirocha} {et~al}\mbox{.}(2017){Mirocha}, {Furlanetto}, \&
  {Sun}}]{Mirocha2017}
{Mirocha} J., {Furlanetto} S.~R., {Sun} G., 2017, \mnras, 464, 1365

\bibitem[{{Mirocha} {et~al}\mbox{.}(2020){Mirocha}, {Mason}, \&
  {Stark}}]{Mirocha2020}
{Mirocha} J., {Mason} C., {Stark} D.~P., 2020, \mnras, 498, 2645

\bibitem[{Morishita {et~al}\mbox{.}(2018)Morishita, Trenti, Stiavelli, Bradley,
  Coe, Oesch, Mason, Bridge, Holwerda, Livermore, Salmon, Schmidt, Shull, \&
  Treu}]{Morishita2018}
Morishita T. {et~al.}, 2018, \apj, 867, 150

\bibitem[{Murray {et~al}\mbox{.}(2005)Murray, Quataert, \&
  Thompson}]{Murray2005}
Murray N., Quataert E., Thompson T.~A., 2005, \apj, 618, 569

\bibitem[{{Murray} {et~al}\mbox{.}(2013){Murray}, {Power}, \&
  {Robotham}}]{Murray2013}
{Murray} S.~G., {Power} C., {Robotham} A.~S.~G., 2013, Astronomy and Computing,
  3, 23

\bibitem[{{Mutch} {et~al}\mbox{.}(2016){Mutch}, {Geil}, {Poole}, {Angel},
  {Duffy}, {Mesinger}, \& {Wyithe}}]{Mutch2016}
{Mutch} S.~J. {et~al.}, 2016, \mnras, 462, 250

\bibitem[{{Narayanan} {et~al}\mbox{.}(2018){Narayanan}, {Conroy}, {Dav{\'e}},
  {Johnson}, \& {Popping}}]{Narayanan2018}
{Narayanan} D. {et~al.}, 2018, \apj, 869, 70

\bibitem[{{Oesch} {et~al}\mbox{.}(2018){Oesch}, {Bouwens}, {Illingworth},
  {Labb{\'e}}, \& {Stefanon}}]{Oesch2018}
{Oesch} P.~A. {et~al.}, 2018, \apj, 855, 105

\bibitem[{{Oke} \& {Gunn}(1983)}]{Oke1983}
{Oke} J.~B., {Gunn} J.~E., 1983, \apj, 266, 713

\bibitem[{{Orr} {et~al}\mbox{.}(2019){Orr}, {Hayward}, \& {Hopkins}}]{Orr2018}
{Orr} M.~E., {Hayward} C.~C., {Hopkins} P.~F., 2019, \mnras, 486, 4724

\bibitem[{Park {et~al}\mbox{.}(2017)Park, Kim, Liu, Trenti, Duffy, Geil, Mutch,
  Poole, Mesinger, \& Wyithe}]{Park2017}
Park J. {et~al.}, 2017, \mnras, 472, 1995

\bibitem[{{Planck Collaboration} {et~al}\mbox{.}(2018){Planck Collaboration},
  {Aghanim}, {Akrami}, {Ashdown}, {Aumont}, {Baccigalupi}, {Ballardini},
  {Banday}, {Barreiro}, {Bartolo}, {Basak}, {Battye}, {Benabed}, {Bernard},
  {Bersanelli}, {Bielewicz}, {Bock}, {Bond}, {Borrill}, {Bouchet}, {Boulanger},
  {Bucher}, {Burigana}, {Butler}, {Calabrese}, {Cardoso}, {Carron},
  {Challinor}, {Chiang}, {Chluba}, {Colombo}, {Combet}, {Contreras}, {Crill},
  {Cuttaia}, {de Bernardis}, {de Zotti}, {Delabrouille}, {Delouis}, {Di
  Valentino}, {Diego}, {Dor{\'e}}, {Douspis}, {Ducout}, {Dupac}, {Dusini},
  {Efstathiou}, {Elsner}, {En{\ss}lin}, {Eriksen}, {Fantaye}, {Farhang},
  {Fergusson}, {Fernandez-Cobos}, {Finelli}, {Forastieri}, {Frailis},
  {Fraisse}, {Franceschi}, {Frolov}, {Galeotta}, {Galli}, {Ganga},
  {G{\'e}nova-Santos}, {Gerbino}, {Ghosh}, {Gonz{\'a}lez-Nuevo}, {G{\'o}rski},
  {Gratton}, {Gruppuso}, {Gudmundsson}, {Hamann}, {Handley}, {Hansen},
  {Herranz}, {Hildebrandt}, {Hivon}, {Huang}, {Jaffe}, {Jones}, {Karakci},
  {Keih{\"a}nen}, {Keskitalo}, {Kiiveri}, {Kim}, {Kisner}, {Knox},
  {Krachmalnicoff}, {Kunz}, {Kurki-Suonio}, {Lagache}, {Lamarre}, {Lasenby},
  {Lattanzi}, {Lawrence}, {Le Jeune}, {Lemos}, {Lesgourgues}, {Levrier},
  {Lewis}, {Liguori}, {Lilje}, {Lilley}, {Lindholm}, {L{\'o}pez-Caniego},
  {Lubin}, {Ma}, {Mac{\'\i}as-P{\'e}rez}, {Maggio}, {Maino}, {Mandolesi},
  {Mangilli}, {Marcos-Caballero}, {Maris}, {Martin}, {Martinelli},
  {Mart{\'\i}nez-Gonz{\'a}lez}, {Matarrese}, {Mauri}, {McEwen}, {Meinhold},
  {Melchiorri}, {Mennella}, {Migliaccio}, {Millea}, {Mitra},
  {Miville-Desch{\^e}nes}, {Molinari}, {Montier}, {Morgante}, {Moss}, {Natoli},
  {N{\o}rgaard-Nielsen}, {Pagano}, {Paoletti}, {Partridge}, {Patanchon},
  {Peiris}, {Perrotta}, {Pettorino}, {Piacentini}, {Polastri}, {Polenta},
  {Puget}, {Rachen}, {Reinecke}, {Remazeilles}, {Renzi}, {Rocha}, {Rosset},
  {Roudier}, {Rubi{\~n}o-Mart{\'\i}n}, {Ruiz-Granados}, {Salvati}, {Sandri},
  {Savelainen}, {Scott}, {Shellard}, {Sirignano}, {Sirri}, {Spencer},
  {Sunyaev}, {Suur-Uski}, {Tauber}, {Tavagnacco}, {Tenti}, {Toffolatti},
  {Tomasi}, {Trombetti}, {Valenziano}, {Valiviita}, {Van Tent}, {Vibert},
  {Vielva}, {Villa}, {Vittorio}, {Wand elt}, {Wehus}, {White}, {White},
  {Zacchei}, \& {Zonca}}]{Planck2018}
{Planck Collaboration} {et~al.}, 2018, arXiv e-prints, arXiv:1807.06209

\bibitem[{{Popping} {et~al}\mbox{.}(2017){Popping}, {Somerville}, \&
  {Galametz}}]{Popping2017}
{Popping} G., {Somerville} R.~S., {Galametz} M., 2017, \mnras, 471, 3152

\bibitem[{{Qiu} {et~al}\mbox{.}(2019){Qiu}, {Mutch}, {da Cunha}, {Poole}, \&
  {Wyithe}}]{Qiu2019}
{Qiu} Y. {et~al.}, 2019, \mnras, 489, 1357

\bibitem[{{Robertson}(2010)}]{Robertson2010}
{Robertson} B.~E., 2010, \apjl, 716, L229

\bibitem[{{Rojas-Ruiz} {et~al}\mbox{.}(2020){Rojas-Ruiz}, {Finkelstein},
  {Bagley}, {Stevans}, {Finkelstein}, {Larson}, {Mechtley}, \&
  {Diekmann}}]{RojasRuiz2020}
{Rojas-Ruiz} S. {et~al.}, 2020, \apj, 891, 146

\bibitem[{Semenov {et~al}\mbox{.}(2018)Semenov, Kravtsov, \&
  Gnedin}]{Semenov2018}
Semenov V.~A., Kravtsov A.~V., Gnedin N.~Y., 2018, \apj, 861, 4

\bibitem[{Smit {et~al}\mbox{.}(2016)Smit, Bouwens, Labbe, Franx, Wilkins, \&
  Oesch}]{Smit2016}
Smit R. {et~al.}, 2016, \apj, 833, 254

\bibitem[{Somerville {et~al}\mbox{.}(2012)Somerville, Gilmore, Primack, \&
  Dom{\'\i}nguez}]{Somerville2012}
Somerville R.~S. {et~al.}, 2012, \mnras, 423, 1992

\bibitem[{{Song} {et~al}\mbox{.}(2016){Song}, {Finkelstein}, {Ashby},
  {Grazian}, {Lu}, {Papovich}, {Salmon}, {Somerville}, {Dickinson}, {Duncan},
  {Faber}, {Fazio}, {Ferguson}, {Fontana}, {Guo}, {Hathi}, {Lee}, {Merlin}, \&
  {Willner}}]{Song2016}
{Song} M. {et~al.}, 2016, \apj, 825, 5

\bibitem[{{Stefanon} {et~al}\mbox{.}(2017){Stefanon}, {Bouwens}, {Labb{\'e}},
  {Muzzin}, {Marchesini}, {Oesch}, \& {Gonzalez}}]{Stefanon2017}
{Stefanon} M. {et~al.}, 2017, \apj, 843, 36

\bibitem[{Stefanon {et~al}\mbox{.}(2019)Stefanon, Labbe, Bouwens, Oesch, Ashby,
  Caputi, Franx, Fynbo, Illingworth, Le~Fevre, Marchesini, McCracken,
  Milvang-Jensen, Muzzin, \& van Dokkum}]{Stefanon2019}
Stefanon M. {et~al.}, 2019, \apj, 883, 99

\bibitem[{{Sun} \& {Furlanetto}(2016)}]{Sun2016}
{Sun} G., {Furlanetto} S.~R., 2016, \mnras

\bibitem[{{Tacchella} {et~al}\mbox{.}(2018){Tacchella}, {Bose}, {Conroy},
  {Eisenstein}, \& {Johnson}}]{Tacchella2018}
{Tacchella} S. {et~al.}, 2018, \apj, 868, 92

\bibitem[{Tacchella {et~al}\mbox{.}(2013)Tacchella, Trenti, \&
  Carollo}]{Tacchella2013}
Tacchella S., Trenti M., Carollo C.~M., 2013, \apjl, 768, L37

\bibitem[{Thompson \& Krumholz(2016)}]{Thompson2016}
Thompson T.~A., Krumholz M.~R., 2016, \mnras, 455, 334

\bibitem[{{Tinker} {et~al}\mbox{.}(2010){Tinker}, {Robertson}, {Kravtsov},
  {Klypin}, {Warren}, {Yepes}, \& {Gottl{\"o}ber}}]{Tinker2010}
{Tinker} J.~L. {et~al.}, 2010, \apj, 724, 878

\bibitem[{Trenti {et~al}\mbox{.}(2010)Trenti, Stiavelli, Bouwens, Oesch, Shull,
  Illingworth, Bradley, \& Carollo}]{Trenti2010}
Trenti M. {et~al.}, 2010, \apj, 714, L202

\bibitem[{{Van Der Walt} {et~al}\mbox{.}(2011){Van Der Walt}, {Colbert}, \&
  {Varoquaux}}]{numpy}
{Van Der Walt} S., {Colbert} S.~C., {Varoquaux} G., 2011, {Computing in Science
  \& Engineering}, 13, 22

\bibitem[{Virtanen {et~al}\mbox{.}(2020)Virtanen, Gommers, Oliphant, Haberland,
  Reddy, Cournapeau, Burovski, Peterson, Weckesser, Bright, {van der Walt},
  Brett, Wilson, Millman, Mayorov, Nelson, Jones, Kern, Larson, Carey, Polat,
  Feng, Moore, {VanderPlas}, Laxalde, Perktold, Cimrman, Henriksen, Quintero,
  Harris, Archibald, Ribeiro, Pedregosa, {van Mulbregt}, \& {SciPy 1.0
  Contributors}}]{scipy}
Virtanen P. {et~al.}, 2020, Nature Methods, 17, 261

\bibitem[{{Vogelsberger} {et~al}\mbox{.}(2020){Vogelsberger}, {Nelson},
  {Pillepich}, {Shen}, {Marinacci}, {Springel}, {Pakmor}, {Tacchella},
  {Weinberger}, {Torrey}, \& {Hernquist}}]{Vogelsberger2019}
{Vogelsberger} M. {et~al.}, 2020, \mnras, 492, 5167

\bibitem[{{Weisz} {et~al}\mbox{.}(2012){Weisz}, {Johnson}, {Johnson},
  {Skillman}, {Lee}, {Kennicutt}, {Calzetti}, {van Zee}, {Bothwell},
  {Dalcanton}, {Dale}, \& {Williams}}]{Weisz2012}
{Weisz} D.~R. {et~al.}, 2012, \apj, 744, 44

\bibitem[{Williams {et~al}\mbox{.}(2018)Williams, Curtis-Lake, Hainline,
  Chevallard, Robertson, Charlot, Endsley, Stark, Willmer, Alberts, Amorin,
  Arribas, Baum, Bunker, Carniani, Crandall, Egami, Eisenstein, Ferruit,
  Husemann, Maseda, Maiolino, Rawle, Rieke, Smit, Tacchella, \&
  Willott}]{Williams2018}
Williams C.~C. {et~al.}, 2018, The Astrophysical Journal Supplement Series,
  236, 33

\bibitem[{{Windhorst} {et~al}\mbox{.}(2011){Windhorst}, {Cohen}, {Hathi},
  {McCarthy}, {Ryan}, {Yan}, {Baldry}, {Driver}, {Frogel}, {Hill}, {Kelvin},
  {Koekemoer}, {Mechtley}, {O'Connell}, {Robotham}, {Rutkowski}, {Seibert},
  {Straughn}, {Tuffs}, {Balick}, {Bond}, {Bushouse}, {Calzetti}, {Crockett},
  {Disney}, {Dopita}, {Hall}, {Holtzman}, {Kaviraj}, {Kimble}, {MacKenty},
  {Mutchler}, {Paresce}, {Saha}, {Silk}, {Trauger}, {Walker}, {Whitmore}, \&
  {Young}}]{Windhorst2011}
{Windhorst} R.~A. {et~al.}, 2011, \apjs, 193, 27

\bibitem[{{Wyithe} {et~al}\mbox{.}(2014){Wyithe}, {Loeb}, \&
  {Oesch}}]{Wyithe2014}
{Wyithe} J. S.~B., {Loeb} A., {Oesch} P.~A., 2014, \mnras, 439, 1326

\bibitem[{{Yung} {et~al}\mbox{.}(2019){Yung}, {Somerville}, {Finkelstein},
  {Popping}, \& {Dav{\'e}}}]{Yung2019a}
{Yung} L.~Y.~A. {et~al.}, 2019, \mnras, 483, 2983

\end{thebibliography}
\bibliographystyle{mn2e_short}

%%%%%%%%%%%%%%%%%%%%%%%%%%%%%%%
%%%%%%%%%%%%%%%%%%%%%%%%%%%%%%%
% APPENDIX
%%%%%%%%%%%%%%%%%%%%%%%%%%%%%%%%
%%%%%%%%%%%%%%%%%%%%%%%%%%%%%%%%
\appendix

%\section{Duty Cycle Methodology}
%Talk about how we used a random duty cycle, but that ``coherent'' burst of star formation, perehaps with some power spectrum of fluctuations, could look different.
%\begin{figure}
%\begin{center}
%\includegraphics[width=0.49\textwidth]{fduty_colours.pdf}
%\caption{{\bf Burstiness affects the luminosity and colour of galaxies, and so must be included}}
%\label{fig:fduty_colours}
%\end{center}
%\end{figure}

\end{document}